\documentclass[pre,twocolumn,groupedaddress,superscriptaddress]{revtex4}

\usepackage[latin1]{inputenc}
\usepackage{amsmath}
\usepackage{hyperref}
\usepackage{amsfonts}
\usepackage{amssymb}
\usepackage{graphicx}
\usepackage{float}
\usepackage{subfigure}
\usepackage[ugly]{units}
\usepackage[T1]{fontenc}
\usepackage{url} 
\usepackage{lmodern}
\usepackage{setspace}  

\usepackage{booktabs} 

\makeatletter
\renewcommand\@biblabel[1]{#1.} 
\makeatother
\bibpunct{(}{)}{,}{n}{,}{,}

\begin{document}

\begin{center}
  \large{TITLE PAGE}
\end{center}
\noindent\textbf{Full title:} \textit{Fractal scale-invariant and
  nonlinear properties of cardiac dynamics remain stable with
  advanced age: A new mechanistic picture of cardiac control in healthy elderly}
\\~\\
\textbf{Running title:} \textit{Fractal and nonlinear stability of cardiac dynamics with aging}
\\
\begin{singlespace}
\noindent
\textbf{Authors:}\\
Daniel T. Schmitt$^{1,2}$, 
Plamen Ch. Ivanov$^{2,3}$, 
\\

\noindent$^1$ Theoretische Physik\\
Universität Ulm\\
D-89069 Ulm\\
Germany\\
$^2$ Center for Polymer Studies and Department of Physics\\
Boston University\\
Boston, MA, 02215\\
USA\\
$^3$ Harvard Medical School and Devision of Sleep Medicine\\
Brigham and Women's Hospital\\
Boston, MA, 02115\\
USA\\
\\
\textbf{Please, direct correspondence to:}\\
Daniel T. Schmitt and  Plamen Ch. Ivanov\\
Center for Polymer Studies and Department of Physics\\
Boston University\\
Boston, MA, 02215\\
Tel: +1 (617) 353-4518\\
FAX: +1 (617) 353-9393\\
dtsch@buphy.bu.edu\\
plamen@buphy.bu.edu\\
\end{singlespace}

\thispagestyle{empty} 
\newpage{}
\begin{abstract}
  \noindent{}\textbf{Background--} Heartbeat
  fluctuations exhibit temporal structure with robust long-range
  correlations, fractal and nonlinear features, which have been found
  to break down with pathologic conditions, reflecting changes in the
  mechanism of neuroautonomic control. It has been hypothesized that
  these features change and even break down also with advanced age,
  suggesting fundamental alterations in cardiac control with aging.
  Here we test this hypothesis.  \\\textbf{Methods and Results--} We
  analyze heartbeat interval recordings from two independent
  databases: (a) 19 healthy young (avg. age $25.7$ years) and 16
  healthy elderly subjects (avg. age $73.8$ years) during 2h under
  resting conditions from the Fantasia database; and (b) 29 healthy
  elderly subjects (avg. age $75.9$ years) during $\approx{}8$h of
  sleep from the SHHS database, and the same subjects recorded 5 years
  later.  We quantify: (1) The average heart rate $\left<RR\right>$;
  (2) the SD $\sigma_{RR}$ and $\sigma_{\Delta{}RR}$ of the heartbeat
  intervals $RR$ and their increments $\Delta{}RR$; (3) the long-range
  correlations in $RR$ as measured by the scaling exponent
  $\alpha_{RR}$ using the Detrended Fluctuation Analysis; (4) fractal
  linear and nonlinear properties as represented by the scaling
  exponents $\alpha^{\mathrm{sign}}$ and $\alpha^{\mathrm{mag}}$ for
  the time series of the sign and magnitude of $\Delta{}RR$; (5) the
  nonlinear fractal dimension $D(k)$ of $RR$ using the Fractal
  Dimension Analysis. We find: (1) No significant difference in
  $\left<RR\right>$ ($P>0.05$); (2) a significant difference in
  $\sigma_{RR}$ and $\sigma_{\Delta{}RR}$ for the Fantasia groups
  ($P<10^{-4}$) but no significant change with age between the elderly SHHS
  groups ($P>0.5$); (3) no significant change in the fractal measures
  $\alpha_{RR}$ ($P>0.15$), $\alpha^{\mathrm{sign}}$ ($P>0.2$),
  $\alpha^{\mathrm{mag}}$ ($P>0.3$), and $D(k)$ with age.
  \\\textbf{Conclusions--} Our findings do not support the hypothesis
  that fractal linear and nonlinear characteristics of heartbeat
  dynamics break down with advanced age in healthy subjects.  While
  our results indeed show a reduced SD of heartbeat fluctuations with
  advanced age, the inherent temporal fractal and nonlinear organization of
  these fluctuations remains stable. This indicates that the coupled
  cascade of nonlinear feedback loops, which are believed to underlie
  cardiac neuroautonomic regulation, remains intact with advanced age.
\\~
\\
\noindent{}\textbf{Keywords:} aging; dynamics; heart rate; nervous system, autonomic; physiology; sleep; fractals; nonlinearity; scaling
\end{abstract}

\newpage{}
The outputs of physiologic systems under neural regulation exhibit (a)
high degree of variability, (b) spacial and temporal fractal
organization which remains invariant at different scales of
observation, as well as (c) complex nonlinear
properties~\cite{Book:Kitney1980,
  Book:Bassingthwaighte1994, Book:Malik1995}. These inherent features
of physiologic dynamics change significantly with different
physiologic states such as wake and sleep, exercise and rest,
circadian rhythms, as well as with pathologic conditions
. As different physiological states and pathologic perturbations
correspond to changes or even breakdown in the mechanism of the
underlying neural regulation, alterations in certain dynamical
properties of physiologic signals have been found to be reliable
markers of changes in physiologic control
.

Aging is traditionally associated with the process of decline of
physiologic function and reduction of physiologic
complexity~\cite{Journal:Kaplan1991, Book:Arking2006}. One major
hypothesis is that physiologic aging results from a gradual change in
the underlying mechanisms of physiologic control --- a regulatory
network of neural and metabolic pathways interacting through coupled
cascades of nonlinear feedback loops on a range of time and length
scales --- leading to changes of physiologic dynamics. Under this
hypothesis even ostensibly healthy elderly subjects would exhibit: (i)
Loss of sensitivity and decreased 
responsiveness to external and internal stimuli, leading to reduced
physiologic variability~\cite{Book:Arking2006}; (ii) Breakdown of
certain feedback loops acting at different time scales in the
regulatory mechanism of various physiologic systems. This breakdown
would lead to loss of physiologic complexity as reflected in certain
scale-invariant and nonlinear temporal characteristics of physiologic
dynamics~\cite{Journal:Kaplan1991, Journal:Lipsitz1992Jama}. This
hypothesis of a breakdown of physiologic complexity with healthy aging
has recently been challenged~\cite{Journal:Vaillancourt2002}. Further,
earlier studies have linked various pathologic states with breakdown
of the scale-invariant fractal organization in physiologic dynamics,
which is likely to result from disintegration of coupled feedback
loops in the regulatory mechanism~\cite{Journal:Saul1988,
  Journal:Glass1990,Journal:Peng1993PRL, Journal:Yamamoto1995,
  Journal:Ivanov1998EurophysL, Journal:Ivanov2001Chaos}. Thus, based
on this hypothesis, mechanistically, physiologic processes under
healthy aging would be categorized in the same class as pathologic
dynamics where fractal organization and nonlinear complexity is lost.

A second hypothesis is that while aging may lead to reduced
variability, certain temporal fractal, scale-invariant and nonlinear
structures embedded in physiologic dynamics may remain unchanged.
These two alternative hypotheses represent different notions about
which aspects of the physiologic control mechanisms are expected to
change in the process of aging in contrast to the changes accompanying
certain pathologic conditions.

To test these two hypotheses we analyze cardiac dynamics --- a typical
example of an output of an integrated physiologic system under
autonomic neural regulation. Previous studies have shown that heart
rate variability decreases with certain pathologic
conditions~\cite{Journal:Wolf1978
  , Book:Malik1995}, as well as with advanced
age~\cite{Journal:OBrien1986, Journal:Tsuji1994Circulation}. Studies
based on approaches from statistical physics and nonlinear dynamics
revealed that heartbeat fluctuations in healthy subjects possess a
self-similar fractal structure characterized by long-range power-law
correlations over a range of time scales~\cite{Journal:Kobayashi1982,
  Journal:Saul1988, Journal:Peng1993PRL}. The scaling exponent
associated with these power-law correlations was shown to change
significantly with rest and exercise~\cite{Journal:Karasik2002PRE,
  Journal:Martinis2004, Journal:Echeverria2006},
posture~\cite{Journal:Butler1993, Journal:Yeragani1994,
  Journal:Tulppo2001}, sleep and wake
state~\cite{Journal:Ivanov1999EPL}, across sleep
stages~\cite{Journal:Bunde2000PRL, Journal:Kantelhardt2002PRE,
  Journal:Penzel2003IEEE,Journal:Kantelhardt2003EPL,
  Journal:Penzel2003Neuropsychopharmacology,
  Journal:Staudacher2005PhysicaA} and circadian
phases~\cite{Journal:Huikuri1990Cardiol,
  Journal:Molgaard1991,Journal:Hu2004PNAS}, and to be a reliable
marker of cardiac vulnerability under pathologic
conditions~\cite{Journal:Peng1995Chaos, Journal:Bigger1996Circulation,
  Journal:Ho1997Circulation, Journal:Amaral1998PRL,
  Journal:Havlin1999PhysicaA_App}.  Further, studies have found that
turbulence-like multifractal and nonlinear features in heartbeat
dynamics are reduced and even lost with
disease~\cite{Journal:Ivanov1999Nature, Journal:Stanley1999PhysicaA,
  Journal:Ivanov2001Chaos, Journal:Lin2001}. Several studies have also
reported reduced heart rate
variability~\cite{Journal:Umetani1998JAmCollCardiol} (as also shown in
Fig.~\ref{fig:timeseries_fantasia}), apparent loss of fractal
organization, as well as breakdown of scale-invariant correlations and
certain nonlinear properties with advanced
age~\cite{Journal:Lipsitz1992Jama, Journal:Iyengar1996,
  Journal:Pikkujamsa1999Circulation, Journal:Vargas2003PRE,
  Journal:Goldberger2002Commentary}, suggesting that healthy aging is
associated with changes in the neuroautonomic mechanism of cardiac
regulation related to disintegration of coupled feedback loops across
a range of time scales.

Here, we investigate how cardiac dynamics change with advanced age by
analyzing scale-invariant, linear, and nonlinear characteristics of
heartbeat fluctuations recorded from subjects during rest and sleep
from two independent databases.

\section{Data and Methods}\label{sec:data}
We analyze heartbeat interval recordings from two independent
databases.

\begin{figure}[htbp]
  \centering 
    \includegraphics[width=0.95\linewidth]{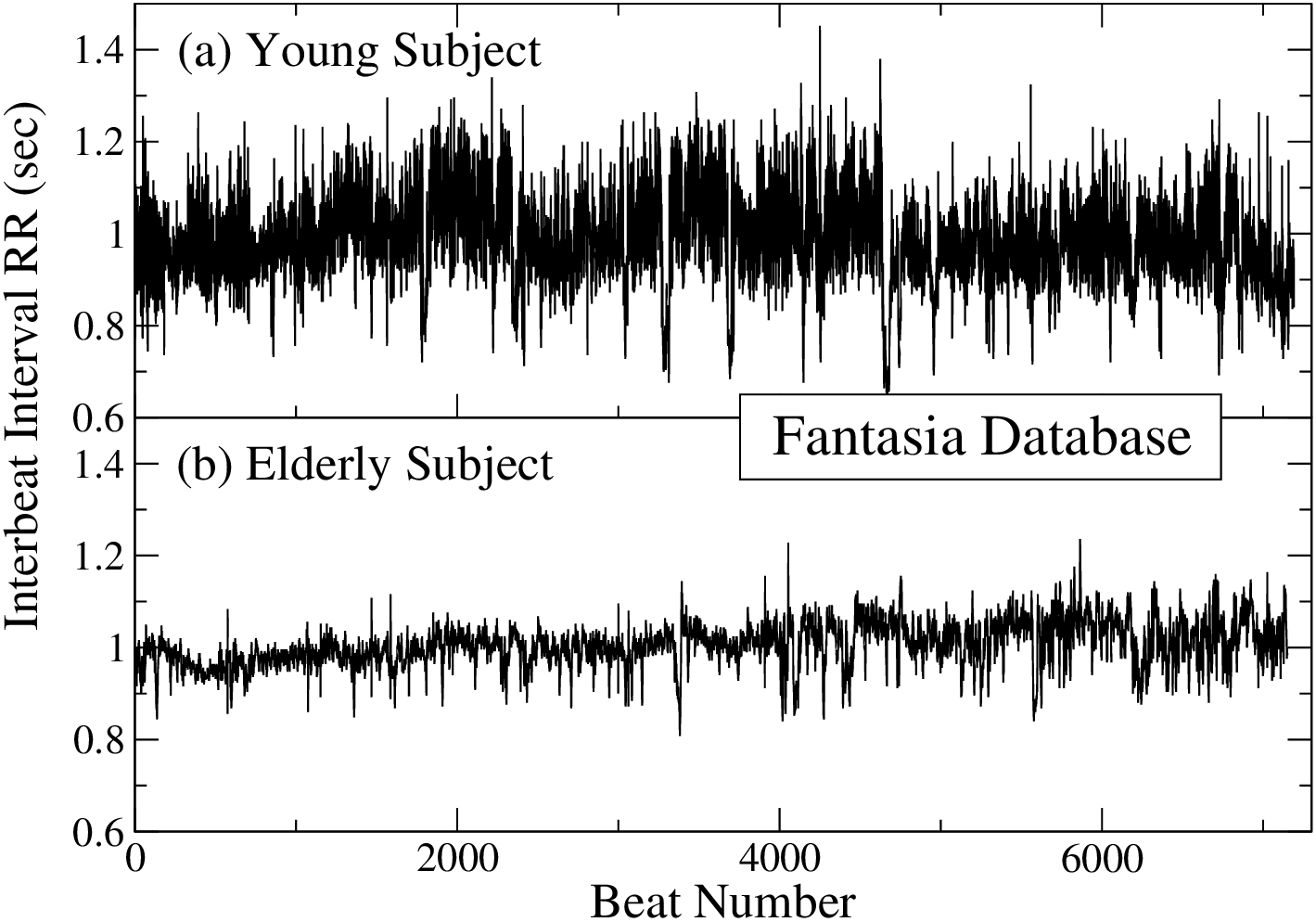}
    \caption{Consecutive heartbeat RR intervals from a representative
      (a) young healthy and (b) elderly healthy subject from the Fantasia
      database. Under the same resting conditions elderly subjects
      exhibit significantly reduced heart rate variability.}
  \label{fig:timeseries_fantasia}
\end{figure}

\emph{Fantasia Database:} The Fantasia database~\cite{PhysioNet}
contains $20$ young and $20$ elderly subjects.  We carefully selected $19$
healthy young subjects ($9$ male; $10$ female) with an average age of
\unit[25.7]{years} (youngest $21$; oldest $34$) and $16$ healthy elderly
subjects ($6$ male; $10$ female) with an average age \unit[73.8]{years}
(youngest $68$; oldest $85$).  All subjects were recorded while watching
the movie Fantasia (Disney, $1940$) in a relaxed supine or
semi-recumbent posture.  These conditions were chosen to avoid the
effect which differences in the level of physical activity between
young and elderly subjects during daily routine might have on cardiac
dynamics (Fig.~\ref{fig:timeseries_fantasia}).  The continuous ECG and
respiration signals were digitized at \unit[250]{Hz}.  Each heartbeat
was annotated using the ARISTOTLE arrhythmia
detector~\cite{Journal:Aristotle}, and each beat annotation was
verified by visual inspection. Only intervals between two normal beats
were considered. One young and four elderly subjects (shown in
Fig.~\ref{fig:excluded}) were excluded from our analysis due to
artefacts in the data.

\begin{figure}[htbp]
  \centering 
    \includegraphics[width=0.95\linewidth]{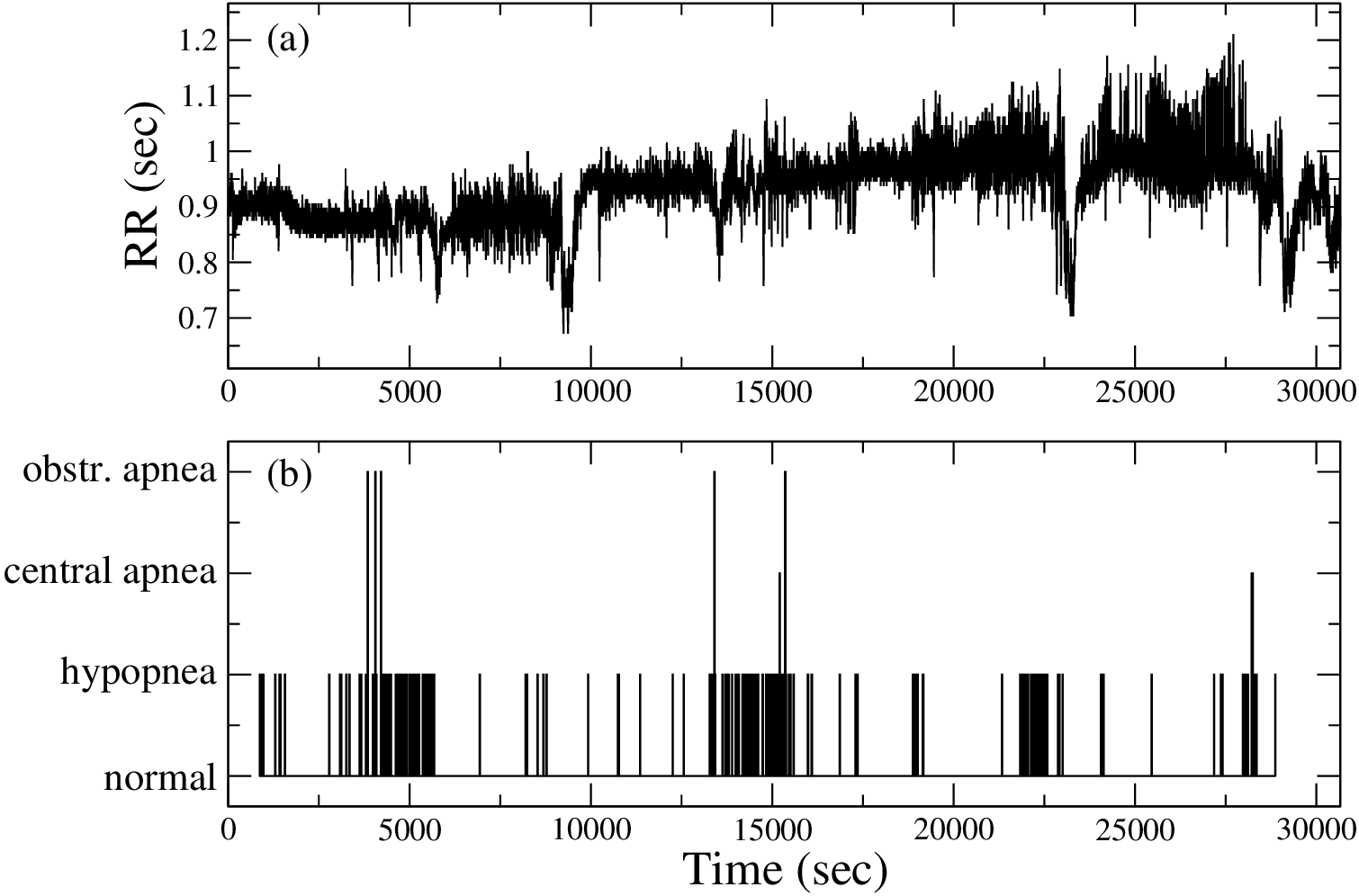}
    \caption{A representative subject from the SHHS database. (a)
      Consecutive heartbeat RR intervals and (b) apnea scoring.}
  \label{fig:timeseries_stein}
\end{figure}

\emph{Sleep Heart Health Study (SHHS) Database:} The Sleep Heart
Health Study (SHHS) is a prospective cohort study designed to
investigate the relationship between sleep disordered breathing and
cardiovascular disease. Subjects were recorded during their habitual
sleep periods of $\approx\unit[8]{h}$, and continuous ECG were
recorded with \unit[250]{Hz} (Fig.~\ref{fig:timeseries_stein}).  Full details
of the study design and cohort are provided
in~\cite{Journal:SHHS_rationale, Journal:SHHS2_methods}. Details about
obtaining the ECG and polysomnographic recordings are outlined
in~\cite{Journal:SHHS_methods}.  Sleep apnea episodes were annotated,
and heart rate data during apnea (obstructive and central) were
excluded from our analysis (Fig.~\ref{fig:timeseries_stein}).  We
selected a subset of 29 subjects (8 males; 21 females) average age at
the time of the first recording \unit[75.9]{years} (youngest 72;
oldest 84). The recordings were repeated \unit[5]{years} later when
the subjects were again screened and categorized as healthy.

\label{sec:methods}
\emph{Detrended Fluctuation Analysis (DFA):}
\label{sec:DFA_method}
We use the DFA method~\cite{Journal:Peng1994}, which has been
developed to quantify fractal correlations embedded in nonstationary
signals, to estimate \textit{dynamic} scale-invariant characteristics
in heartbeat fluctuations.  Compared with traditional correlation
analyzes such as autocorrelation, power-spectrum analysis, and Hurst
analysis, the advantage of the DFA method is that it can accurately
quantify the correlation property of signals masked by polynomial
trends, and is described in detail in~\cite{Journal:Kun2001PRE,
  Journal:Kantelhardt2001PhysicaA, Journal:Chen2002_PRE,
  Journal:xu2005PRE, Journal:Chen2005PRE}.

The DFA method quantifies
the detrended fluctuations $F(n)$ of a signal at different time scales
$n$.  A power-law functional form $F(n)\sim{}n^{\alpha}$ indicates the
presence of self-similar organization in the fluctuations. The
parameter $\alpha$, called the scaling exponent, quantifies the
correlation properties of the heartbeat signal: if $\alpha=0.5$, there
is no correlation and the signal is white noise; if $\alpha=1.5$, the
signal is a random walk (Brownian motion); if $0.5<\alpha<1.5$, there
are positive correlations, where large heartbeat intervals are more
likely to be followed by large intervals (and vice versa for small
heartbeat intervals).

One advantage of the DFA method is that it can quantify signals with
$\alpha>1$, which cannot be done using the traditional autocorrelation
and $R/S$ analyses~\cite{Book:Feder1988}.  In contrast to the
conventional methods, the DFA method avoids spurious detection of
apparent long-range correlations that are an artefact of
nonstationary~\cite{taqqu95}.  Thus, the DFA method is able to detect
subtle temporal structures in highly heterogeneous physiologic time
series.

An inherent limitation of the DFA analysis is the maximum time scale
$n_{\mathrm{max}}$ for which the fluctuation function $F(n)$ can be reliably
calculated. To ensure sufficient statistics at large scales it was
shown that $n_{\mathrm{max}}$ should be chosen
$n_{\mathrm{max}}\le{}N/6$, where $N$ is the length of the
signal~\cite{Journal:Kun2001PRE, Journal:Coronado2005JBioPhys,
  Journal:xu2005PRE}.  For time scales $n<n_{\mathrm{max}}$ there is no bias in
estimating the scaling exponent $\alpha$.  Thus, recordings longer
than $1$ hour ($N\approx 3600$ beats) are sufficient to reliably
quantify $\alpha$ up to time scales $n=600$ beats, and differences in the
length of the recordings between the Fantasia database (2 hours) and
SHHS database (8 hours) do not effect the estimate of $\alpha$.
Recent studies have tested the performance of the DFA method, when
applied to correlated signals with patches of missing data, random
spikes, superposed trends related to different activity levels and
patches with different standard deviation and local correlations, as
often found in heartbeat data~\cite{Journal:Kun2001PRE,
  Journal:Chen2002_PRE}.

Both the Fantasia database and the NIH Sleep Heart Health Study
database have used $250$ Hz sampling rate for the ECG recordings.  A
precision of $0.004$ sec ($250$ Hz) is more than sufficient for our
analysis, since the DFA method as well as the MSA and FDA analyses we
employ (see below) are robust in that respect.  Using a lower sampling
rate (i.e., lower precision in the estimate of the RR-intervals) acts
effectively as added random noise with an amplitude proportional to the
sampling interval --- in our case the amplitude of this ``sampling
noise'' is more than two orders of magnitude smaller than the RR
interval.  It has been shown that adding noise with such a small
amplitude to a fractal correlated signal does not effect the
correlation scaling and fractal
properties~\cite{Journal:Chen2002_PRE}.

\emph{Magnitude and Sign Analyzes (MSA):}
Since the DFA method quantifies \textit{linear} fractal characteristics
related to two-point correlations, we have selected the MSA method to probe
for long-term \textit{non-linear} properties in the data.
Specifically, it has been shown that signals with identical temporal
organization, quantified by the DFA-scaling exponent $\alpha$, can
exhibit very different non-linear properties captured by the MSA
method~\cite{Journal:Ashkenazy2001PRL}.

The MSA method~\cite{yossicc2000, 
  Journal:AshkenazyPhysicaA_2003} consists of the following steps: (i)
given $RR_i$ series we obtain the increment series, $\Delta
RR_i=RR_{i+1}-RR_i$; (ii) we decompose the increment series into a
magnitude series $|\Delta RR|$ and a sign series ${\rm sign}(\Delta
RR)$; (iii) to avoid artificial trends we subtract the average from
the magnitude series; (iv) because of limitations in the accuracy of
the DFA method for estimating the scaling exponents of anti-correlated
signals ($\alpha <0.5$), we integrate the magnitude
series~\cite{Journal:Kun2001PRE}; (v) we perform a scaling analysis
using DFA; (vi) to obtain the scaling exponents for the magnitude
series we measure the slope of $F(n)/n$ on a log-log plot, where
$F(n)$ is the fluctuation function and $n$ is the time scale of
analysis.

This approach is sensitive to nonlinear features in
signals~\cite{Journal:Theiler1992PhysicaD}.  We find that positive
correlations in the magnitude series ($\alpha_{\rm mag}>0.5$) are a
reliable marker of long-term nonlinear properties. Thus, we employ the
MSA as a complementary method to the DFA, because it can
distinguish physiologic signals with identical long-range
correlations, as quantified by the DFA method, but with different nonlinear
properties and different temporal organization for the ${\rm sign}(\Delta
RR)$ series.

\emph{Fractal Dimension Analysis (FDA):}
The fractal dimension $D(k)$ is a \textit{local} nonlinear measure used to
quantify the irregularity of a time series~\cite{Book:Mandelbrot1977}.
We estimate the fractal dimension using an algorithm proposed
in~\cite{Journal:Higuchi_1988}.

Starting from a discrete time series, $x(i)$, with $i
\in [1,N]$, a new sparse time series $x_k^m$ is constructed in the following
way
\begin{equation}
  \label{eq:fractal_dim_sparse_TS}
  x_k^m;\, x\left(m\right), x\left(m+k\right),
  \dots, x\left(m+\left\lfloor{}\frac{N-m}{k}\right\rfloor{} k \right),
\end{equation}
with $m \in \left[1,k\right]$ where $m$ and $k$ are integers, and
$\lfloor{}\frac{N-m}{k}\rfloor{}$ denotes the largest integer number smaller than
$\frac{N-m}{k}$.  Then a length measure for this sparse time series is defined as
\begin{equation}
  \label{eq:fractal_dim_L}
  L_m(k)=\frac{N-1}{h\; k^2} \left(
      \sum_{i=1}^{h} \left|
        x_{ik}^m-x_{(i-1)k}^m \right|
     \right),
\end{equation}
with $h\equiv{}\left\lfloor{}\frac{N-m}{k}\right\rfloor{}$.  For a time
series $x(i)$ with a fractal dimension $D$ the length $L_m(k)$
averaged over $m$ is a power-law function of the scale $k$:
$L(k)\equiv\left<L(k)\right>_m \sim k^{-D}$.  In the general case $D$
can depend on the scale $k$. In this case, the local fractal dimension $D(k)$
of the time series $x(i)$ is defined as the negative local derivative
of $\log L(k)$ as a function of $\log k$.

\begin{table*}[htbp]
  \centering
  \caption{Overview of measures used.}
  \begin{tabular*}{\linewidth}{@{\extracolsep{\fill}}llp{0.4\linewidth}}
    \toprule
    \textbf{Abbreviation} &\multicolumn{1}{c}{\textbf{Measure}}&\textbf{Significance}\\
    \hline
    &\multicolumn{1}{c}{\emph{static measures}} & \\
    $\left<RR\right>$ (AVNN)& mean of $RR$ intervals & inversely
    proportional to heart rate\\
    $\sigma_{\mathrm{RR}}$ (SDNN)& std. deviation of $RR$& para- and sympathetic HRV measure sensitive to trends\\
    $\sigma_{\Delta\mathrm{RR}}$ (RMSSD)& std. deviation of
    $\Delta{}RR$ & parasympathetic HRV measure insensitive to trends\\
    &\multicolumn{1}{c}{\emph{dynamic measures}} & \\
    $\alpha$ & scaling exponent of $RR$& linear scale-invariant correlations\\
    $\alpha^{\mathrm{mag}}$ & scaling exponent of $|\Delta{}RR|$ & nonlinear scale-invariant correlations\\
    $\alpha_{1}^{\mathrm{sgn}}$ & scaling exponent of
    $\mathrm{sgn}(\Delta{}RR)$ & fractal measure of directionality\\
    $D(k)$ & fractal dimension of $RR$ & nonlinear fractal
    measure\\
    \bottomrule
  \end{tabular*}
\label{tab:methods}
\end{table*}

\section{Results}
\subsection{Variability in heartbeat intervals and their increments}
We first test the possibility that advanced age in ostensibly healthy
subjects would lead to an increase in the average heart rate and to a
significant reduction in heart rate variability --- a behavior
previously observed in subjects with congestive heart failure where
under suppressed vagal tone increased heart rate is associated with
reduced heart rate variability~\cite{Journal:Wolf1978,
  Journal:Yamamoto1995, Journal:Goldberger2002PNAS}. We find that both
young and elderly healthy subjects in the Fantasia database exhibit
very similar group average interbeat intervals:
$\left<RR\right>\pm\sigma=0.9\pm0.14$ for the young group and
$\left<RR\right>\pm\sigma=1.06\pm0.17$ for the elderly group, where
$\sigma$ is the standard deviation (Table~\ref{tab:allresults}).  This
is in agreement with previous studies~\cite{Journal:Iyengar1996,
  Journal:Pikkujamsa1999Circulation, Journal:Corino2006}. A Student's
t-test shows no significant difference between the two groups with a
$p\mathrm{-value}=0.11$.  A very similar average heartbeat interval we
observe for the healthy elderly subjects in the SHHS database with
$\left<RR\right>\pm\sigma=0.92\pm0.075$, indicating no significant
difference ($p\mathrm{-value}=0.07$) compared to the group of young
Fantasia subjects (Table~\ref{tab:allresults}).  Further, comparing
the group average heartbeat interval of the elderly subjects from the
SHHS database with the same subjects recorded \unit[5]{years} later,
we find again no significant difference:
$\left<RR\right>\pm\sigma=0.92\pm0.08$ at the first recording and
$\left<RR\right>\pm\sigma=0.92\pm0.1$ after \unit[5]{years}
($p\mathrm{-value}=0.92$, Table~\ref{tab:allresults}).  Thus, we do
not observe a significant change in the average heart rate with
advanced age.

To test whether there is a reduction in heart rate variability with
aging, we next estimate for each subject the standard deviation of the
heartbeat intervals $\sigma_{\mathrm{RR}}$ (often denoted as SDNN) and
the standard deviation of the increments in the consecutive heartbeat
intervals $\sigma_{\Delta\mathrm{RR}}$ (often denoted as RMSSD)
(Table~\ref{tab:allresults}).  For the young and elderly subjects in
the Fantasia database we find a statistically significant difference
with (i) a higher value for the group average
$\left<\sigma_{\mathrm{RR}}\right>$, and (ii) larger inter-subject
variability for the young group:
$\left<\sigma_{\mathrm{RR}}\right>\pm\sigma=0.089\pm0.034$ for the young 
compared to 
$\left<\sigma_{\mathrm{RR}}\right>\pm\sigma=0.051 \pm 0.017$ for the
elderly subjects ($p\mathrm{-value}=3.3\cdot{}10^{-04}$) 
(Table~\ref{tab:allresults}).  Similarly, we observe a significantly
higher value for the group average
$\left<\sigma_{\Delta\mathrm{RR}}\right>$ for the young subjects in
the Fantasia database
($\left<\sigma_{\Delta\mathrm{RR}}\right>\pm\sigma=0.061 \pm 0.031$)
compared to the elderly subjects
($\left<\sigma_{\Delta\mathrm{RR}}\right>\pm\sigma=0.027 \pm 0.012$)
($p\mathrm{-value}=9.9\cdot{}10^{-5}$), again with a larger
inter-subject variability for the young group
(Table~\ref{tab:allresults}).  We note that the sampling rate of
\unit[250]{Hz} does not effect the significance of the difference in
$\sigma_{\Delta\mathrm{RR}}$ between the young and elderly groups, as
this difference is approximately \unit[0.034]{sec}, i.e., one
magnitude larger than the sampling precission of \unit[0.004]{sec}.

For the group of healthy elderly subjects from the SHHS database we
find a higher value of
$\left<\sigma_{\mathrm{RR}}\right>\pm\sigma=0.077\pm 0.027$ compared
to the elderly group from the Fantasia database --- a difference which
could be attributed to the fact that the SHHS subjects were recorded
during sleep where transitions between sleep stages are associated
with trends and larger fluctuations in the interbeat interval time
series~\cite{Journal:Kantelhardt2002PRE,
  Journal:Penzel2003Neuropsychopharmacology, Journal:Penzel2003IEEE},
while the elderly Fantasia subjects were recorded during rest.  In
contrast, for $\left<\sigma_{\Delta\mathrm{RR}}\right>$ we do not
observe a significant difference between the elderly groups from the
Fantasia and SHHS database ($p\mathrm{-value}=0.74$)
(Table.~\ref{tab:allresults}).  However, we find a significant
difference between young and elderly subjects, indicating a clear
reduction in the heart rate variability with aging.

\subsection{Fractal Correlations}
\begin{figure}[htbp]
  \centering 
  \subfigure
  { 
    \label{fig:DFA_F_n_a}
    \includegraphics[width=\linewidth]{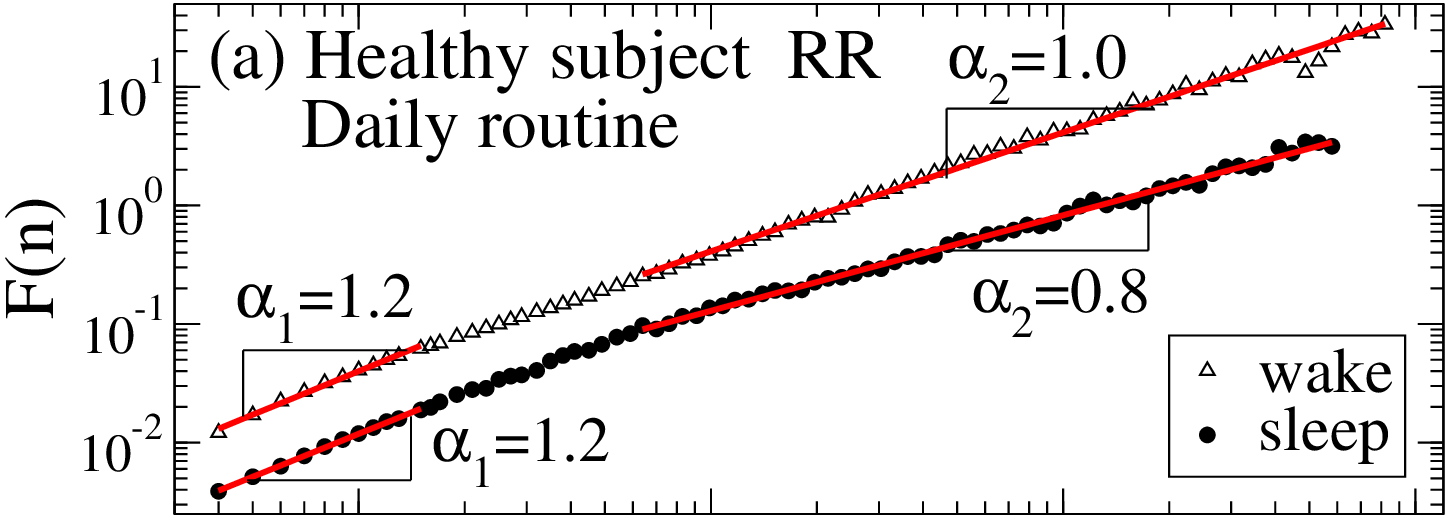}
  }\\
  \vspace{-0.5cm}
  \subfigure
  { 
    \label{fig:DFA_F_n_b}
    \includegraphics[width=\linewidth]{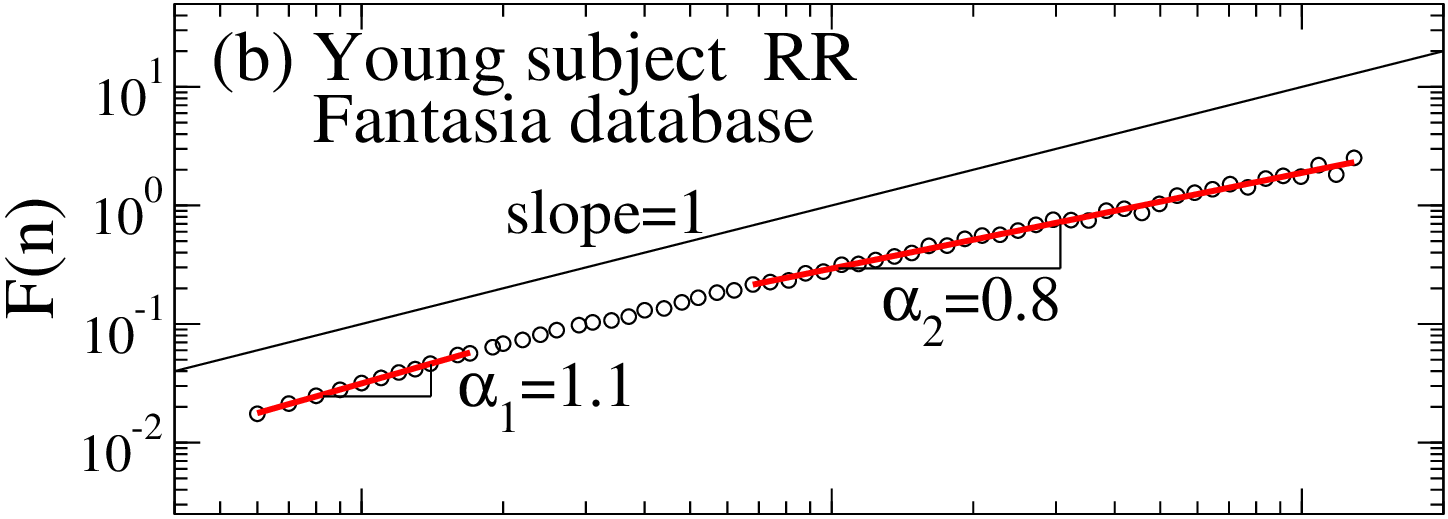}
  }\\
  \vspace{-0.5cm}
  \subfigure
  { 
    \label{fig:DFA_F_n_c}
    \includegraphics[width=\linewidth]{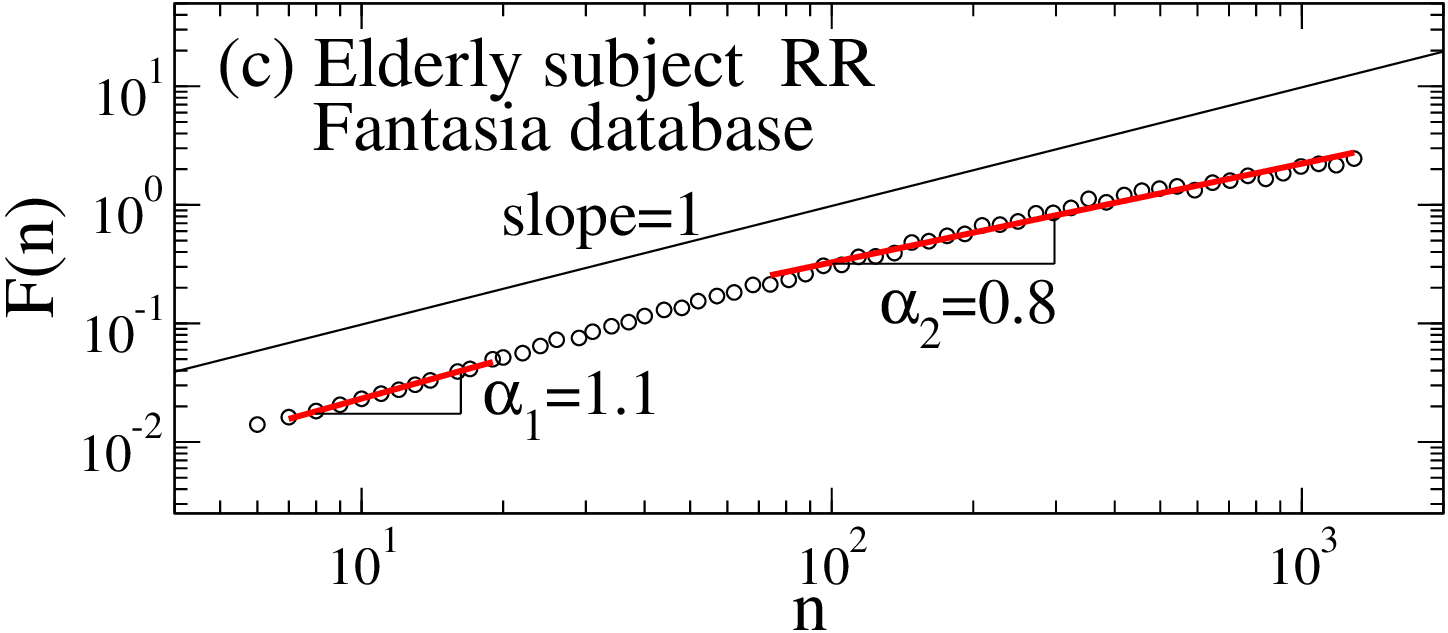}
  }\\
  \vspace{-0.5cm}
  \caption{Fluctuation function $F(n)$ vs.\ time scale $n$ (in heart
    beat number) obtained using \mbox{DFA-$2$} for (a) \unit[6]{h}-long
    record of RR heartbeat intervals during wake and sleep from a
    representative healthy subject (MIT-BIH Normal Sinus Rhythm
    Database~\cite{PhysioNet}), as well as \unit[2]{h}-long records of
    a representative (b) healthy young subject and (c) healthy elderly
    subject from the Fantasia database. A very similar scaling
    behavior is observed for the representative (b) young and (c)
    elderly subjects, which closely resemble the scaling behavior of
    the healthy subjects during sleep shown in (a) (MIT-BIH Normal
    Sinus Rhythm Database~\cite{PhysioNet}), indicating \emph{no change} in
    the scale-invariant temporal correlations of heartbeat intervals
    with advanced age under healthy resting conditions. Scaling curves
    for all individuals are shown in Appendix, Fig.~\ref{fig:Fn_RR_all}.}
  \label{fig:DFA_F_n_all_rep}
\end{figure}

We next test whether the temporal organization in the heartbeat
fluctuations changes in ostensibly healthy elderly compared to young
subjects. Earlier studies have shown that heartbeat fluctuations
exhibit self-similar power-law correlations over a broad range of time
scales ranging from seconds to many hours~\cite{Journal:Kobayashi1982,
  Journal:Saul1988}, and that the scaling exponents associated with
these power-law correlations change significantly with sleep and wake
state~\cite{Journal:Ivanov1999EPL} and with pathologic
conditions~\cite{Journal:Peng1993PRL, Journal:Peng1995Chaos},
reflecting changes in the underlying mechanism of cardiac regulation.
Specifically, heartbeat fluctuations of healthy subjects during daily
activity exhibit $1/f$-like power
spectrum~\cite{Journal:Kobayashi1982, Journal:Saul1988,
  Journal:Peng1993PRL} with a scaling exponent $\alpha\approx{}1$ (see
Section~\ref{sec:methods}). During sleep this behavior changes to
exponent $\alpha\approx{}0.8$ at time scales above $60$ beats,
indicating stronger anti-correlations in the interbeat increments
$\Delta{}$RR during sleep compared to wake
state~\cite{Journal:Ivanov1999EPL} (Fig.~\ref{fig:DFA_F_n_a}). In
contrast, for pathologic conditions such as congestive heart failure
earlier studies have reported a value for the exponent $\alpha$ closer
to $1.5$ --- typical for random walk behavior (Brownian motion) and
associated with loss of cardiac control~\cite{Journal:Peng1995Chaos}.

Applying the DFA method we obtain a very similar scaling behavior for
a representative healthy young and a healthy elderly subject from the
Fantasia database, both characterized by a scaling exponent
$\alpha_2\approx{}0.8$ at intermediate and large time scales (Fig.
\ref{fig:DFA_F_n_b} and \ref{fig:DFA_F_n_c}). At small time scales for
both representative subjects we observe a crossover to a higher
exponent of $\alpha_1\approx{}1.1$ (Fig.~\ref{fig:DFA_F_n_b} and
\ref{fig:DFA_F_n_c}).  While there is certain inter-subject
variability in the scaling functions $F(n)$, this crossover behavior
remains robust with a group average scaling exponent
$\alpha_1\approx{}1.1$ at small scales and $\alpha_2\approx{}0.75$ at
large scales for the young subjects, and respectively
$\alpha_1\approx{}1.2$ and $\alpha_2\approx{}0.8$ for the elderly
subjects (Appendix, Fig.~\ref{fig:Fn_RR_all}). Our analysis indicates no
significant difference in the scaling behavior between healthy young
and healthy elderly subjects under the resting conditions in the
Fantasia study protocol (Table~\ref{tab:scaling_exp_DFA-2}). We note
that our findings for the young and elderly Fantasia subjects (Fig.
\ref{fig:DFA_F_n_b} and \ref{fig:DFA_F_n_c}) are very similar to the
scaling behavior in heartbeat fluctuations previously reported for
healthy subjects during sleep~\cite{Journal:Ivanov1999EPL}, which
exhibit a crossover from $\alpha_1\approx{}1.2$ at small time scales
to $\alpha_2\approx{}0.8$ at intermediate and large time scales (Fig.
\ref{fig:DFA_F_n_a}). This similarity in the scaling properties of
heartbeat dynamics of healthy subjects during sleep (Fig.
\ref{fig:DFA_F_n_a}) and the Fantasia database subjects (Fig.
\ref{fig:DFA_F_n_b} and \ref{fig:DFA_F_n_c}) may be attributed to the
fact that under the Fantasia study protocol subjects are resting in a
semi-recumbent/supine posture, watching a relaxing movie ---
physiologic conditions which more closely resemble sleep than daytime
activity.

\begin{figure}[htbp]
  \centering 
    \includegraphics[width=\linewidth]{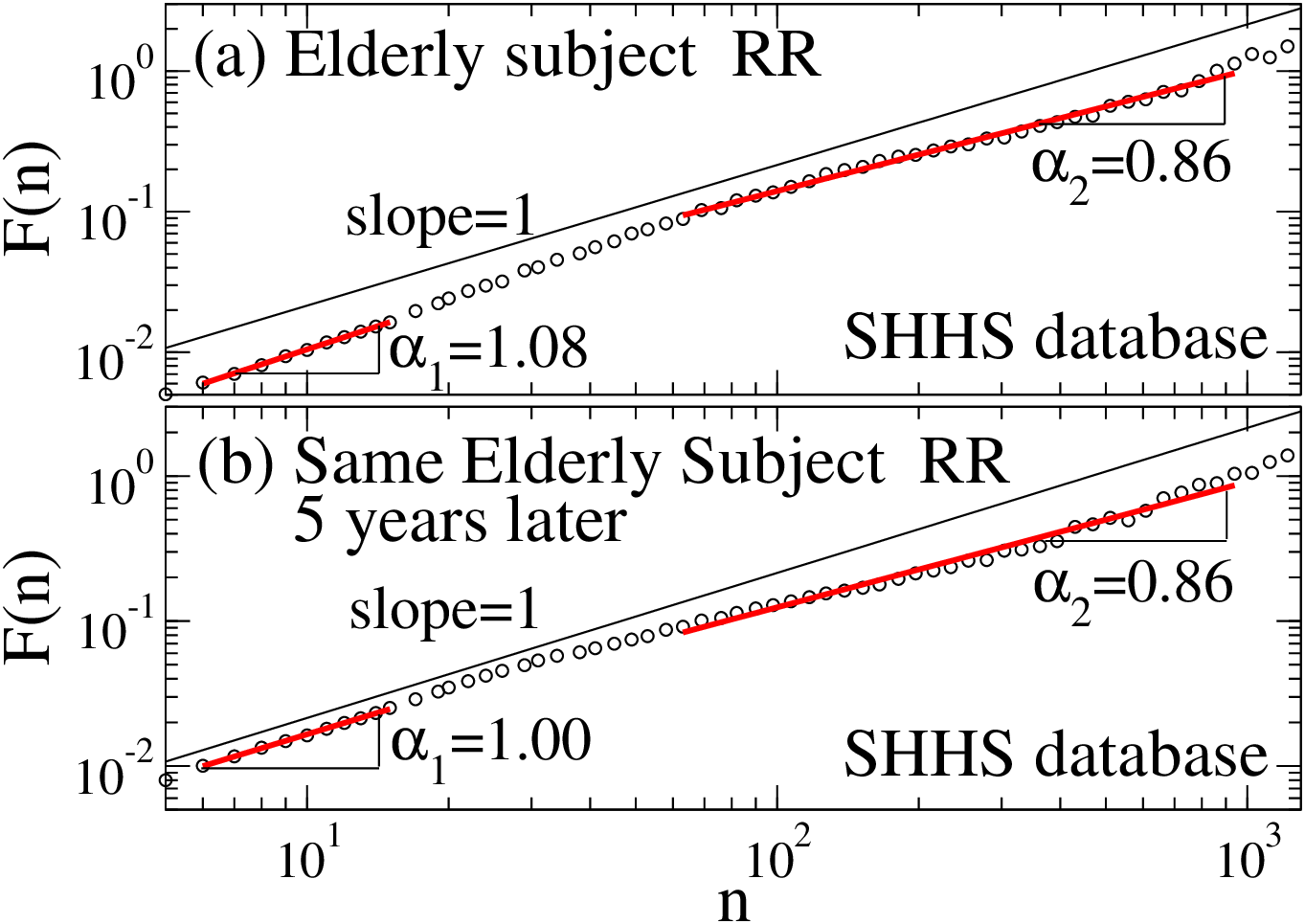}
    \caption{Fluctuation function $F(n)$ vs. time scale $n$ (in heart
      beat number) obtained from \mbox{DFA-$2$} for
      $\approx{}\unit[8]{h}$ long records of RR heartbeat intervals
      during sleep for (a) representative healthy elderly subject from
      the SHHS database and (b) the same elderly subject
      \unit[5]{years} later. The very similar values for the exponents
      $\alpha_1$ and $\alpha_2$ indicate no breakdown of linear
      fractal correlations with advanced age under healthy conditions.
      Note the similarity with the scaling behavior for the young
      subjects, shown in Fig.~\ref{fig:DFA_F_n_all_rep} and Appendix,
      Fig.~\ref{fig:Fn_RR_all}, which is \emph{not} consistent with
      the hypothesis of a gradual loss of scale-invariant complexity
      in the process of aging.}
  \label{fig:Fn_rr_stein_rep}
\end{figure}

To confirm the validity of these findings, we further investigate the
scale-invariant correlation properties of cardiac dynamics for healthy
elderly subjects from the SHHS database, where heart rate data were
recorded during sleep --- a protocol which differs from the Fantasia
study (see Section~\ref{sec:data}). In Fig.~\ref{fig:Fn_rr_stein_rep}
we show the DFA scaling curves for a representative SHHS subject with
a crossover in the scaling behavior from $\alpha_1\approx{}1.1$ at
small time scales to $\alpha_2\approx{}0.9$ above \unit[60]{beats}.
This scaling behavior is very similar to the one we find for both
young and elderly subjects from the Fantasia database (Fig.
\ref{fig:DFA_F_n_all_rep}). Further, comparing the scaling behavior of
the elderly subjects from the SHHS database to the same subjects
recorded five years later, we do not find a significant difference in
the correlation scaling exponents $\alpha_1$ and $\alpha_2$
(Fig.~\ref{fig:Fn_rr_stein_rep} and Table~\ref{tab:stein_scaling_exp_DFA-2}). The results shown in Figs.
\ref{fig:DFA_F_n_all_rep}, \ref{fig:Fn_rr_stein_rep} and in Appendix,
Fig.~\ref{fig:Fn_RR_all} indicate that the fractal correlation
properties of healthy heartbeat dynamics remain stable and do not
significantly change with advanced age.

\subsection{Magnitude and sign scaling analysis (MSA)}
\label{sec:results_MSA}
Recent studies have demonstrated that scale-invariant processes with
identical long-range power-law correlations may be characterized by
very different dynamics for the magnitude and sign of their
fluctuations~\cite{Journal:Ashkenazy2001PRL,
  Journal:Kantelhardt2002PRE}, and that the information contained in
the temporal organization of the magnitude and the sign time series is
independent from the correlation properties of the original time
series~\cite{Journal:AshkenazyPhysicaA_2003}. Specifically, for
cardiac dynamics of healthy subjects it was
shown~\cite{Journal:Ashkenazy2001PRL} that heartbeat intervals during
routine daily activity exhibit correlation properties at intermediate
and large time scales characterized by scaling exponent
$\alpha_2\approx{}1$, while at the same time scales the magnitude
series of the increments in consecutive heartbeat intervals is
characterized by $\alpha_2^{\mathrm{mag}}\approx{}0.8$. Further, while
correlations reflect the linear properties of heartbeat dynamics, the
temporal structure of the magnitude of interbeat increments has been
shown to relate to the nonlinear properties encoded in the Fourier
phases~\cite{Journal:Ashkenazy2001PRL, Journal:AshkenazyPhysicaA_2003,
  Journal:Theiler1992PhysicaD}.  For certain pathologic conditions
such as congestive heart failure previous studies have reported loss
of nonlinearity~\cite{Journal:Poon1997Nature}
associated with a
breakdown of the multifractal
spectrum~\cite{Journal:Ivanov1999Nature}, and reduced scaling exponent
$\alpha^{\mathrm{mag}}$ for the magnitude
series~\cite{Journal:AshkenazyPhysicaA_2003}.

\begin{figure}[htbp]
  \centering
  \subfigure{
    \includegraphics[width=\linewidth]{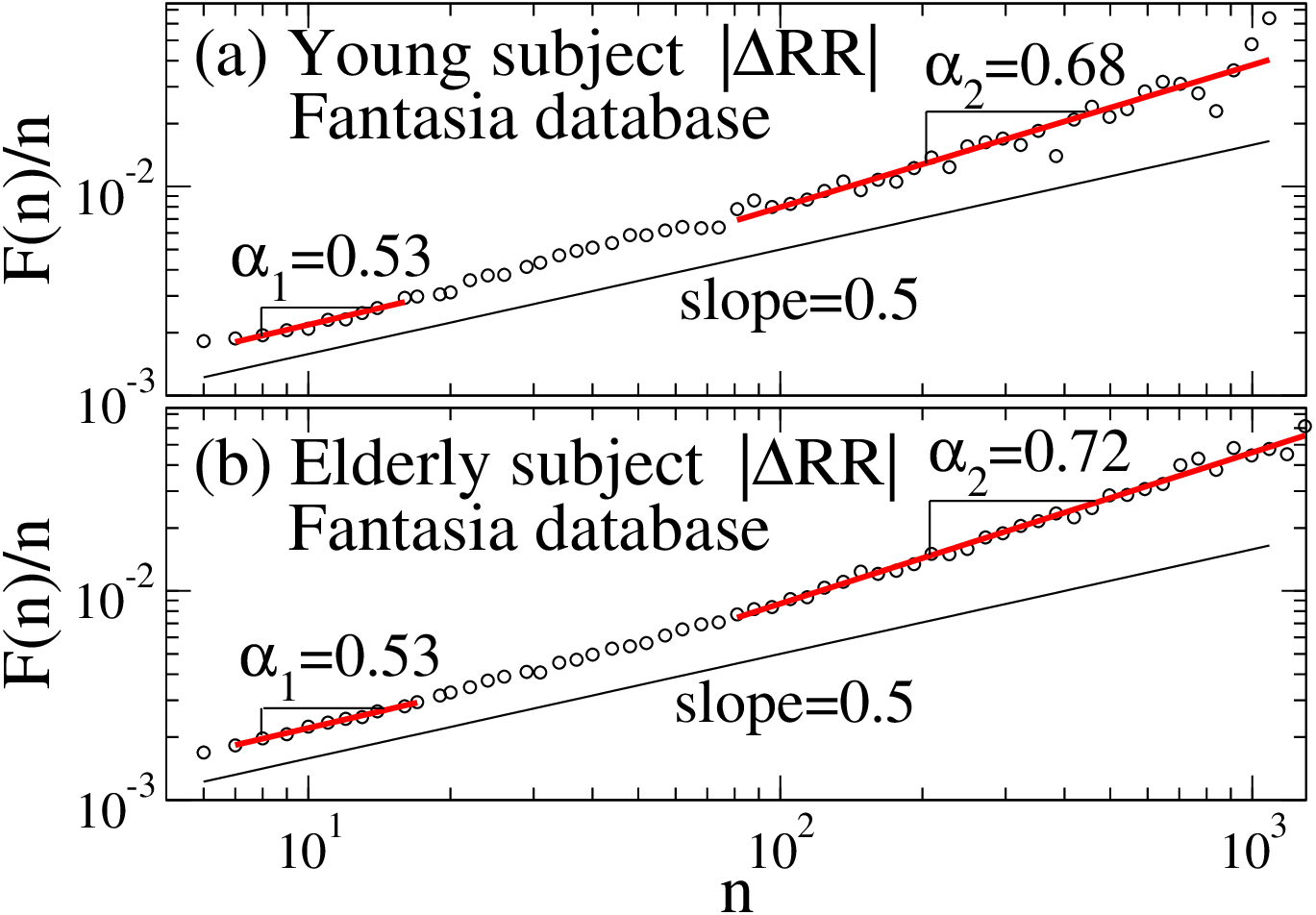}
    \label{fig:Fn_abs}
  }\\
  \vspace{-0.2cm}
  \subfigure{
    \includegraphics[width=\linewidth]{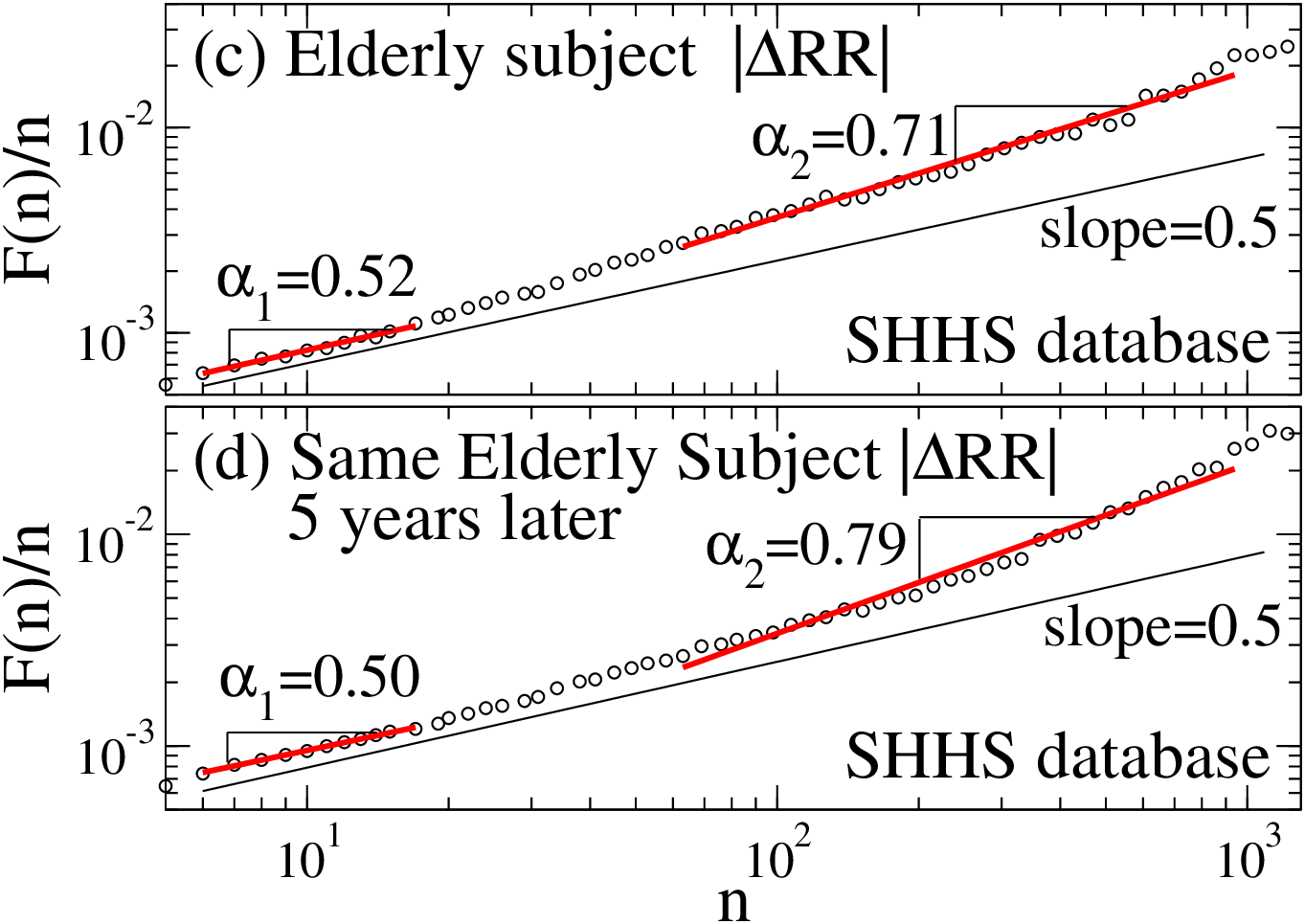}
    \label{fig:Fn_abs_stein_rep}
  }
  \vspace{-0.2cm}
  \caption{Fluctuation function $F(n)$ vs. time scale $n$ (in beat
    number) obtained for the magnitude of the interbeat
    increments $|\Delta{}RR|$ using \mbox{DFA-$2$} for a representative (a)
    healthy young and (b) healthy elderly subject from the Fantasia
    database, and for a representative (c) healthy elderly subject
    from the SHHS database and (d) the same subject recorded
    \unit[5]{years} later.  All subjects exhibit a very similar
    scaling behavior characterized by an exponent
    $\alpha_2^{\mathrm{mag}}\approx{}0.7$ at intermediate and large
    time scales, very different than $\alpha^{\mathrm{mag}}=0.5$
    characteristic for linear processes with no correlations in the
    Fourier phases~\cite{Journal:Ashkenazy2001PRL,
      Journal:AshkenazyPhysicaA_2003}, which indicates that the
    long-term nonlinear properties of heartbeat dynamics do not break
    down with advanced age under healthy resting conditions. This is
    in contrast to the hypothesis linking the process of healthy aging
    with a gradual loss of nonlinearity. Scaling curves for all
    individuals from the Fantasia database are shown in Appendix,
    Fig.~\ref{fig:Fn_abs_young_old}.}
  \label{fig:Fn_abs_all}
\end{figure}

For the magnitude time series of the interbeat increments we obtain
$\alpha_1^{\mathrm{mag}}\approx{}0.53$ at
small time scales and $\alpha_2^{\mathrm{mag}}\approx{}0.68$ at
intermediate and large time scales for a representative young subject
(Fig.~\ref{fig:Fn_abs_all}(a)), and very similar results with
$\alpha_1^{\mathrm{mag}}\approx{}0.53$ and
$\alpha_2^{\mathrm{mag}}\approx{}0.72$ for a representative elderly
subject from the Fantasia database (Fig.~\ref{fig:Fn_abs_all}(b)). The
DFA scaling functions $F(n)$ for all young and elderly subjects, shown in
Appendix, Fig.~\ref{fig:Fn_abs_young_old}, exhibit a
consistent behavior among the subjects in each group with a smooth
crossover from a group average magnitude exponent
$\alpha_1^{\mathrm{mag}}\approx{}0.53$ at small and intermediate time
scales to $\alpha_2^{\mathrm{mag}}\approx{}0.64$ at large scales for
the young group, and a similar crossover from a group average exponent
$\alpha_1^{\mathrm{mag}}\approx{}0.6$ at small and intermediate time
scales to $\alpha_2^{\mathrm{mag}}\approx{}0.7$ at large scales for
the elderly group (Table~\ref{tab:stein_scaling_exp_DFA-2}).

To confirm these findings, we next calculate the magnitude scaling
exponent of the interbeat increments for the elderly subjects from the
SHHS database. Again we observe a crossover from
$\alpha_1^{\mathrm{mag}}\approx{}0.52$ at small scales to
$\alpha_2^{\mathrm{mag}}\approx{}0.7$ at large time scales shown in
Fig.~\ref{fig:Fn_abs_all}(c) for a representative elderly subject ---
a behavior very similar to the one observed for both young and elderly
Fantasia subjects shown in Fig.~\ref{fig:Fn_abs_all}(a--b). Our
analysis does not show a statistically significant difference in the
group average magnitude scaling exponents $\alpha_1^{\mathrm{mag}}$
(with $p$-value=0.71) and $\alpha_2^{\mathrm{mag}}$ (with
$p$-value=0.57) between the elderly SHHS subjects and the elderly
Fantasia subjects.  Moreover, we find no significant difference in
$\alpha_1^{\mathrm{mag}}$ (with $p$-value=0.24) and
$\alpha_2^{\mathrm{mag}}$ (with $p$-value=0.16) between the elderly
SHHS subjects and the young Fantasia subjects.

\begin{figure}[htbp]
  \centering 
  \subfigure{
    \includegraphics[width=\linewidth]{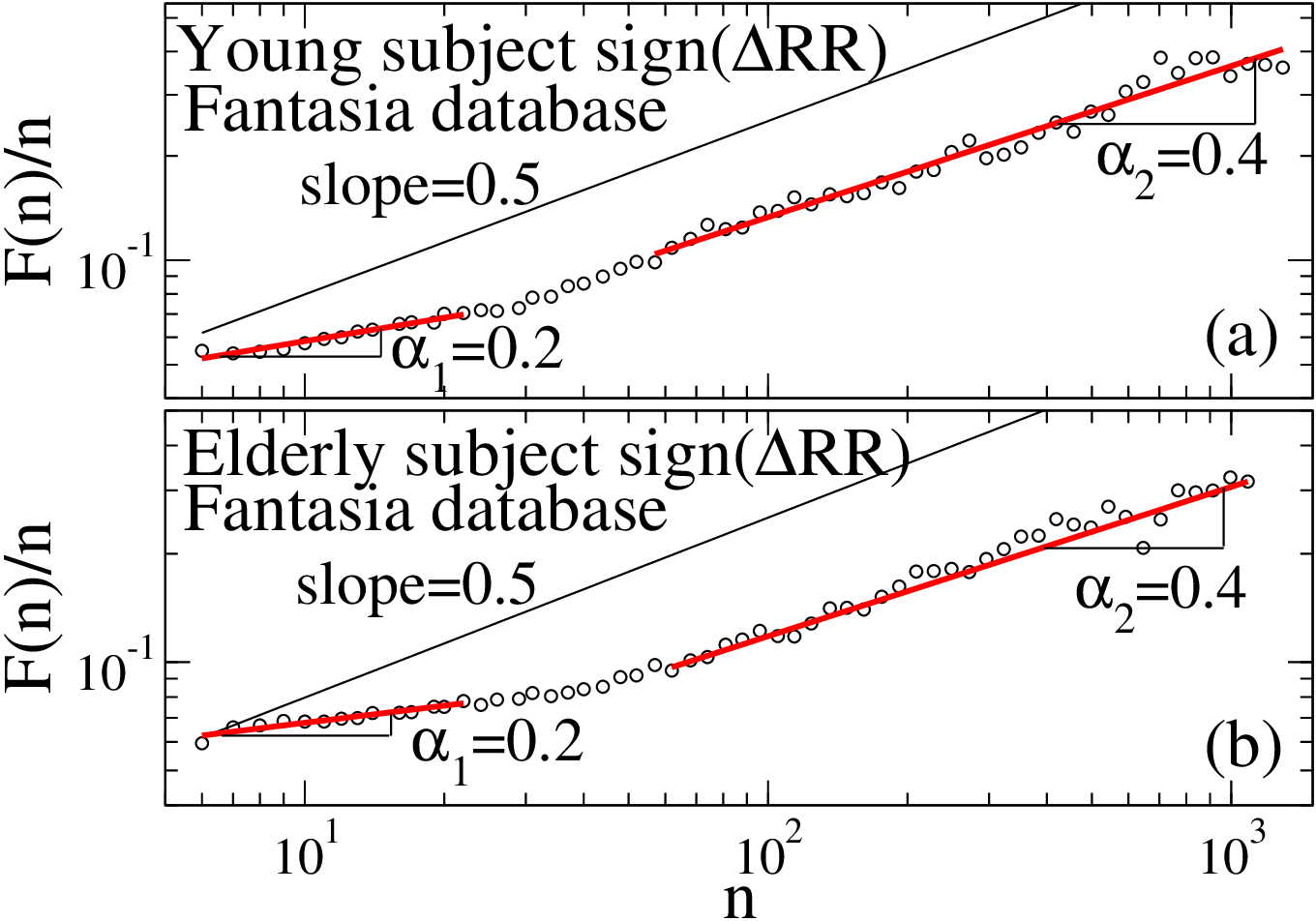}
    \label{fig:Fn_sgn}
  }\\
  \vspace{-0.2cm}
  \subfigure{
    \includegraphics[width=\linewidth]{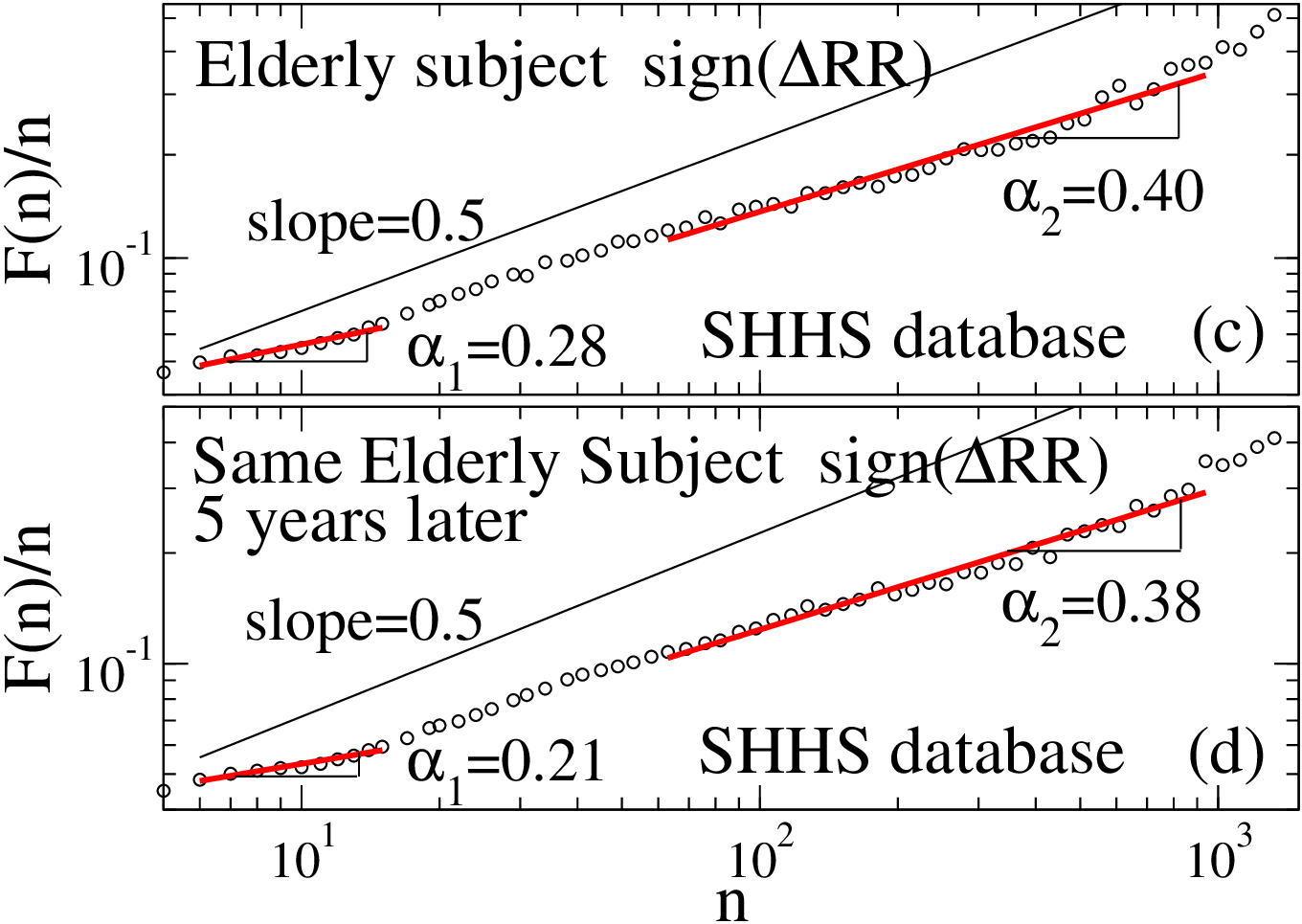}
    \label{fig:Fn_sign_stein_rep}
  }
  \vspace{-0.2cm}
  \caption{Fluctuation function $F(n)$ vs. time scale $n$ (in beat
    numbers) obtained for the sign of the interbeat
    increments $\mathrm{sign}(\Delta{}RR)$ using \mbox{DFA-$2$} for a
    representative (a) healthy young and (b) healthy elderly subject
    in the Fantasia database, and for (c) representative healthy
    elderly subject from the SHHS database and (d) the same elderly
    SHHS subject recorded \unit[5]{years} later. All subjects exhibit
    a very similar scaling behavior for the sign with a crossover from
    strong anti-correlations with
    $\alpha_1^{\mathrm{sgn}}\approx{}0.2$ at small time scales to
    weaker anti-correlations with
    $\alpha_2^{\mathrm{sgn}}\approx{}0.4$ at large scales, indicating
    a similar fractal organization of sympathetic and parasympathetic
    control in both young and elderly subjects under healthy resting
    conditions. Scaling curves for all individuals in the Fantasia
    database are shown in Appendix, Fig.~\ref{fig:Fn_sgn_young_old_all}.
  }
  \label{fig:Fn_sgn_rep_fantasia_SHHS}
\end{figure}

For the sign of the interbeat increments time series we again find
\emph{no} significant difference in the scaling behavior between the
young and elderly subjects in the Fantasia database with practically
identical exponents of $\alpha_1^{\mathrm{sgn}}\approx{}0.2$ at short
time scales and $\alpha_2^{\mathrm{sgn}}\approx{}0.4$ at intermediate
and large time scales (Fig.~\ref{fig:Fn_sgn_rep_fantasia_SHHS}(a--b)).
A consistently similar behavior we observe for all subjects in the
young and elderly group in the Fantasia database (Appendix,
Fig.~\ref{fig:Fn_sgn_young_old_all}), where the scaling function
$F(n)$ exhibits a crossover from strongly anti-correlated behavior at
short time scales to weaker anti-correlations at larger scales,
respectively characterized by group average sign exponents
$\alpha_1^{\mathrm{sgn}}\approx{}0.24$ for the young and
$\alpha_1^{\mathrm{sgn}}\approx{}0.3$ for the elderly subjects at
small scales, and $\alpha_2^{\mathrm{sgn}}\approx{}0.47$ for the young
and $\alpha_2^{\mathrm{sgn}}\approx{}0.43$ for the elderly subjects at
large scales. These results indicate no significant difference in the
temporal organization of the sign series between the young and the
elderly subjects in the Fantasia database
(Table~\ref{tab:scaling_exp_DFA-2}).

Repeating our sign scaling analysis for the SHHS database we observe a
crossover from strongly anti-correlated behavior with an exponent
$\alpha_1^{\mathrm{sgn}}\approx{}0.2$ at small time scales to weaker
anti-correlations with $\alpha_2^{\mathrm{sgn}}\approx{}0.4$ at
intermediate and large time scales, as shown in Fig.
\ref{fig:Fn_sgn_rep_fantasia_SHHS}(c--d).  This crossover behavior is
very similar to the one we find for both young and elderly Fantasia
subjects (Fig.  \ref{fig:Fn_sgn_rep_fantasia_SHHS}(a--b) and Appendix,
Fig.~\ref{fig:Fn_sgn_young_old_all}). Moreover, we do not find a
significant difference in the scaling of the sign series for the
elderly SHHS subjects and the same subjects five years later (Fig.
\ref{fig:Fn_sgn_rep_fantasia_SHHS}(c--d) and
Table~\ref{tab:stein_scaling_exp_DFA-2}).

\subsection{Fractal Dimension Analysis}
\begin{figure}[htbp]
  \centering
  \subfigure
      {
        \includegraphics[width=\linewidth]{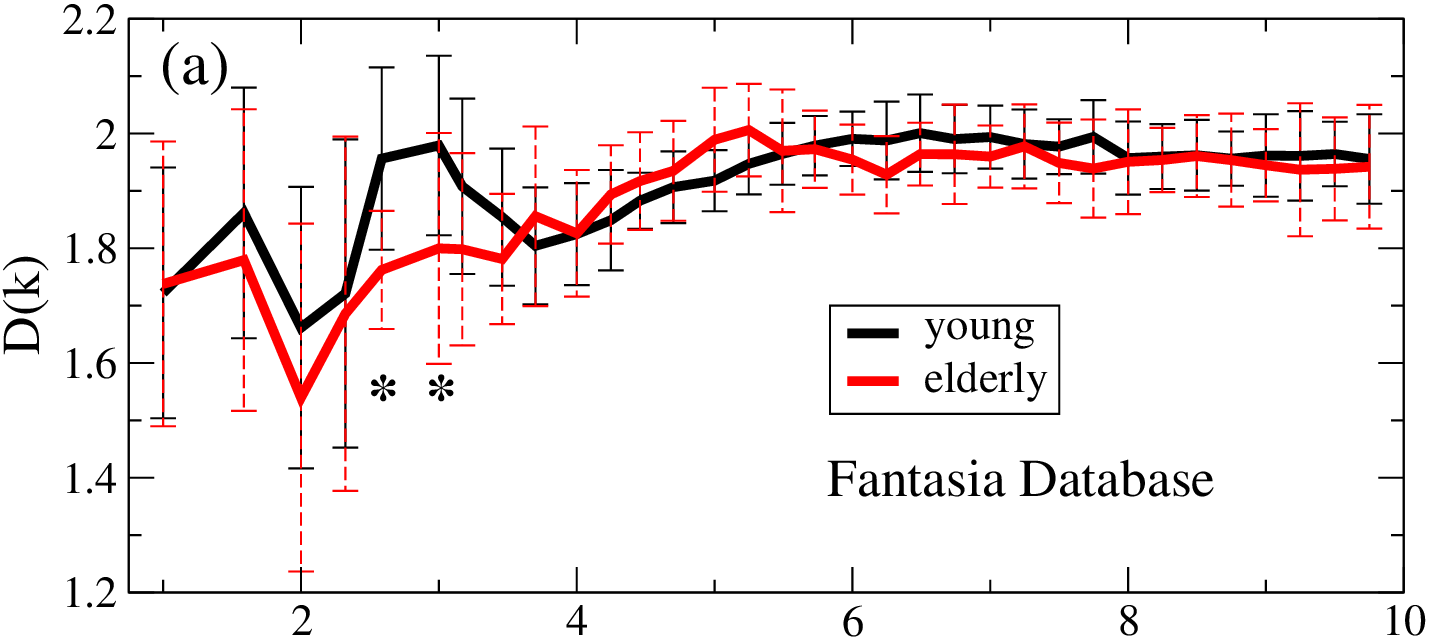}
        \label{fig:FractalDim_fantasia}
      }\\
  \vspace{-0.4cm}
  \subfigure
      {
        \includegraphics[width=\linewidth]{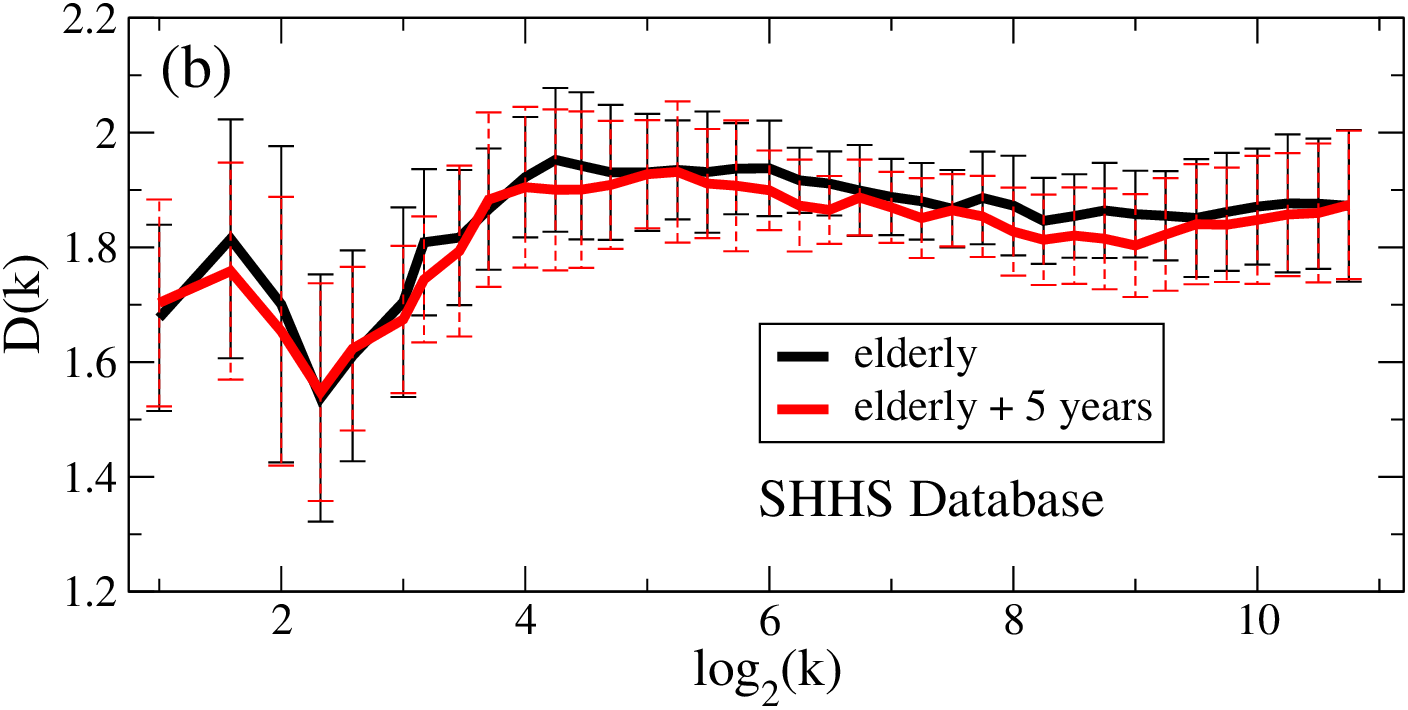}
        \label{fig:FractalDim_stein}
      }
      \caption{Group average nonlinear fractal dimension $D(k)$ versus
        time scale $\log_2 k$, where $k$ is measured in beat numbers.
        (a) For young and elderly healthy Fantasia subjects; (b)
        healthy elderly SHHS subjects and the same subjects
        \unit[5]{years} later. There is no significant difference in
        the group averages indicated by the overlapping standard
        deviations except for the interval of scales $k \in [3,6]$
        marked in (a) by the symbol ($\ast$)
        ($p\mathrm{-value}=1.94\times 10^{-4}$ and $6\times 10^{-3}$
        correspondingly). Note the very similar profile of $D(k)$ for
        all groups indicating no apparent loss of nonlinearity with
        aging, in agreement with our findings for the long-term
        nonlinear properties represented by the magnitude exponent
        $\alpha^{\mathrm{mag}}_2$ shown in Fig.
        \ref{fig:Fn_abs_all}.}
\label{fig:FractalDim}
\end{figure}

Finally, we employ the FDA method (see Section~\ref{sec:methods}) to
estimate the fractal dimension $D(k)$ of a time
series~\cite{Book:Mandelbrot1977, Book:Feder1988,
  Journal:Higuchi_1988}.  It has been demonstrated that the fractal
dimension is a measure which represents the nonlinear properties in
the output of a dynamical system, so that two signals with identical
scale-invariant correlations may be quantified by different fractal
dimension depending on the degree of nonlinearity encoded in the
Fourier phases~\cite{Journal:Higuchi_1990,
  Journal:Theiler1992PhysicaD}. Our analysis shows no significant
difference in the group average of the nonlinear fractal dimension
measure $D(k)$ between the young and the elderly subjects in the
Fantasia database for the whole range of time scales except for a very
short time interval of $6$ to $8$ heartbeats (Fig.
\ref{fig:FractalDim_fantasia}) --- time scales typical for sleep apnea
(see Fig.~\ref{fig:excluded}).  At smaller and larger time scales the
average fractal dimension $D(k)$ converges for both groups
(Fig.~\ref{fig:FractalDim_fantasia}).  Furthermore, we do not observe
a statistically significant difference between the elderly subjects
from the SHHS database and the same subjects recorded five years later
(Fig.~\ref{fig:FractalDim_stein}).  These findings do \emph{not}
support the hypothesis that nonlinearity is reduced in healthy elderly
subjects.

\begin{table*}[htbp]
  \centering
  \caption{Average values and standard deviation of 
    $\left<RR\right>$, $\sigma_{\mathrm{RR}}$ (SDNN),
    $\sigma_{\mathrm{\Delta{}RR}}$ (RMSSD), and
    DFA-2 scaling exponents for subjects from the Fantasia database
    and the SHHS database. For the Fantasia database, $F(n)$ was fitted
    in the interval 
    $n \in \left[6,16\right]$ for $\alpha_{1}$ and $n \in
    \left[60,\frac{N}{6}\right]$ for $\alpha_{2}$.  For the SHHS
    database, $F(n)$ was fitted in the interval 
    $n \in \left[6,16\right]$ for $\alpha_{1}$ and $n \in
    \left[60,600\right]$ for $\alpha_{2}$. A two-tailed
      Student's t-test was performed to obtain the p-values.}
  \begin{tabular*}{\linewidth}{@{\extracolsep{\fill}}ccccccc}
    \toprule
    \textbf{} &\multicolumn{3}{c}{\textbf{Fantasia
         Database}}&\multicolumn{3}{c}{\textbf{SHHS Database}}\\
    \cmidrule(r){2-4}\cmidrule(r){5-7}
    Measure & Young & Elderly & $p$-value & Elderly & Elderly + 5y & $p$-value \\
    \hline
     $\left<RR\right>$ & $0.9\phantom{0}\pm0.14$ & $1.06\pm0.17$
    & $0.11$ & $0.92\pm0.08$ & $0.92\pm0.1\phantom{0}$ & $0.92$\\
    
     $\sigma_{\mathrm{RR}}$ & $0.089 \pm 0.034$ & $0.051 \pm 0.017$
    & $\mathbf{3.3\cdot{}10^{-4}}$ & $0.077 \pm 0.027$ & $0.081 \pm 0.024$ & $0.50$\\  
    
     $\sigma_{\Delta\mathrm{RR}}$ & $0.061 \pm 0.031$ & $0.027 \pm 0.012$
    & $\mathbf{9.9\cdot{}10^{-5}}$ & $0.028 \pm 0.015$ & $0.028 \pm 0.013$ & $0.74$\\  
    
    $\alpha_{1}$ & $1.09 \pm 0.24$ & $1.22 \pm 0.29$
    & $0.16$ & $1.12 \pm 0.27$ & $1.09 \pm 0.28$ & $0.78$\\

    $\alpha_{2}$ & $0.76 \pm 0.08 $ & $0.78 \pm 0.12$
    & $0.47$ & $0.88 \pm 0.12$ & $0.97 \pm 0.12$ & $0.01$\\

     $\alpha_{1}^{\mathrm{mag}}$ & $0.53 \pm 0.1\phantom{0}$ & $0.56 \pm 0.08$
    & $0.36$ & $0.57 \pm 0.13$ & $0.60 \pm 0.13$ & $0.49$\\

    $\alpha_{2}^{\mathrm{mag}}$ & $0.64 \pm 0.11$ & $0.68 \pm 0.11$
    & $0.45$ & $0.70 \pm 0.12$ & $0.72 \pm 0.13$ & $0.58$\\

    $\alpha_{1}^{\mathrm{sgn}}$ & $0.24 \pm 0.15$ & $0.3\phantom{0} \pm 0.2\phantom{0}$
    & $0.28$ & $0.23 \pm 0.19$ & $0.21 \pm 0.19$ & $0.74$\\

    $\alpha_{2}^{\mathrm{sgn}}$ & $0.47 \pm 0.09$ & $0.44 \pm 0.08$ & $0.37$
    & $0.38 \pm 0.07$ & $0.39 \pm 0.07$ & $0.77$\\

    \bottomrule
  \end{tabular*}
\label{tab:allresults}
\label{tab:scaling_exp_DFA-2}
\label{tab:stein_scaling_exp_DFA-2}
\end{table*}

\subsection{Summary of the results}
In agreement with previous studies~\cite{Journal:Tsuji1994Circulation,
  Journal:Iyengar1996, Journal:Pikkujamsa1999Circulation,
  Journal:Corino2006} we observe certain degree of reduction in heart
rate variability, as measured by $\sigma_{\mathrm{RR}}$ (SDNN) and
$\sigma_{\mathrm{\Delta{}RR}}$ (RMSSD), when comparing young to
elderly subjects (Table~\ref{tab:allresults}).  In contrast to
previous studies~\cite{Journal:Lipsitz1992Jama, Journal:Iyengar1996,
  Journal:Pikkujamsa1999Circulation}, however, we do \emph{not} find a
significant difference in the scaling exponents $\alpha_1$ and
$\alpha_2$, characterizing the fractal scale-invariant temporal
organization of heartbeat fluctuations, between young and elderly
subjects (Table~\ref{tab:allresults}). For the scaling properties of
the magnitude and the sign of heartbeat fluctuations --- which have
been shown to carry additional independent information about the
nonlinear and linear properties of a time
series~\cite{Journal:Ashkenazy2001PRL, Journal:AshkenazyPhysicaA_2003,
  Journal:Kantelhardt2002PRE} --- we find that these measures also
remain \emph{unchanged} when comparing young and healthy elderly
subjects (Table~\ref{tab:allresults}).  Finally, for the fractal
dimension $D(k)$ of the heartbeat interval time series --- an
independent nonlinear measure --- again contrary to previous
reports~\cite{Journal:Vargas2003PRE}, we do not find significant
differences between young and elderly subjects.  Furthermore,
comparing longitudinal data from a group of elderly subjects who were
also recorded five year later, we find that the heart rate variability
is not further reduced (Table~\ref{tab:allresults}), and that the
scaling exponents $\alpha_1$ and $\alpha_2$ of the heartbeat
fluctuations, as well as the nonlinear features as measured by the
magnitude exponent $\alpha^{\mathrm{mag}}$ and the fractal dimension
$D(k)$, remain stable.

These findings indicate that in the process of aging the alterations
in the underlying mechanisms of cardiac autonomic regulation are not
likely to involve breakdown of coupling between feedback loops at
different time scales or dominance of a particular feedback loop at a
given time scale, as often observed with pathologic
perturbations~\cite{Journal:Kurths1995, Journal:Ivanov1996Nature,
  Journal:Ho1997Circulation, Journal:Maekikallio1997AJC, Journal:Maekikallio1998AJC,
  Journal:Ivanov1998PhysA}.  Rather, our findings suggest a reduced
reflexiveness of the neuroautonomic regulation with aging, while the
nonlinear feedback interactions across time scales between elements of
the cardiac regulatory system remain unchanged.

\section{Interpretation and Modeling}
\begin{figure}[htbp]
  \centering
  \includegraphics[width=\linewidth]{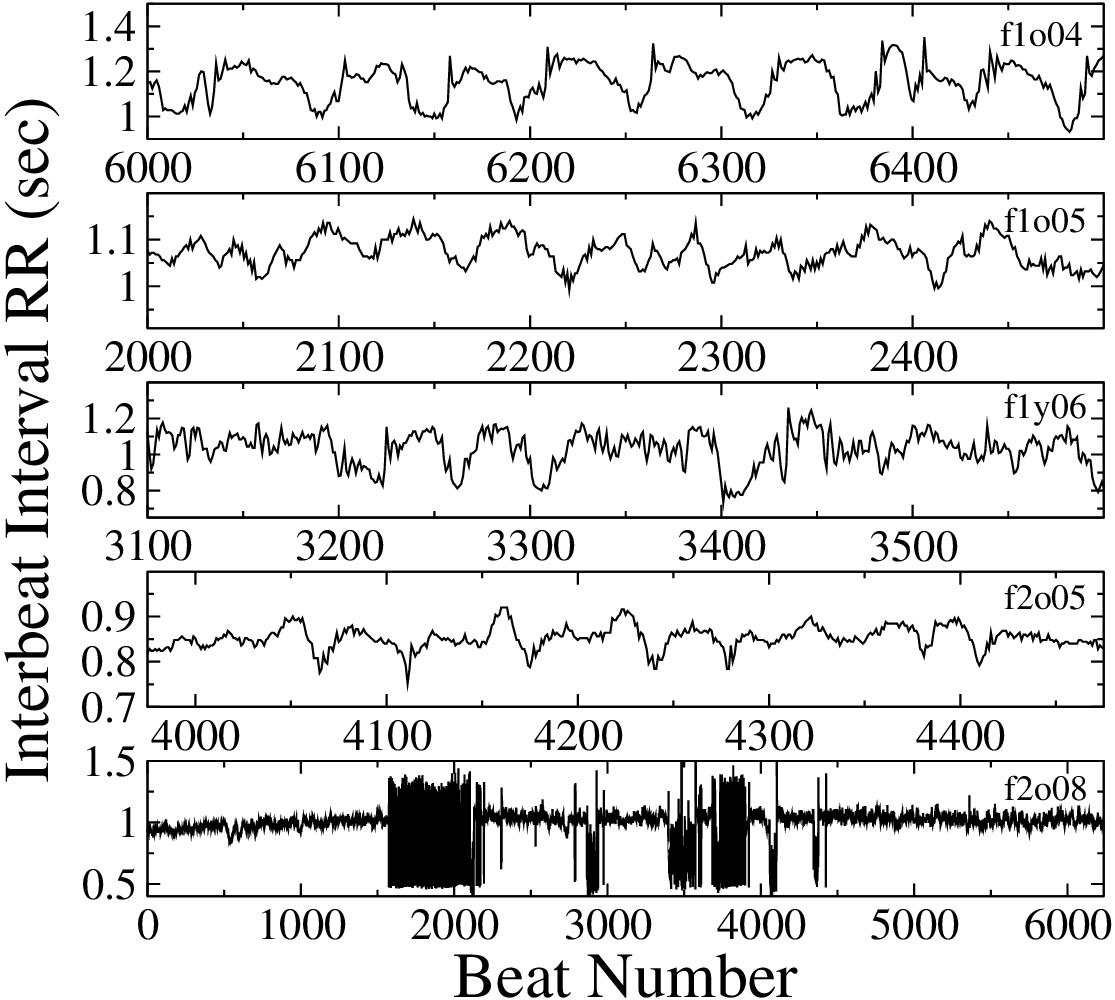} 
  \caption{Recordings of heartbeat intervals from one young and four
    elderly subjects in the Fantasia database excluded from our
    analysis. The first four recordings contain many segments with
    well pronounced periodic patterns, the period and amplitude of
    which are typical for sleep apnea (see Fig.~\ref{fig:apnea_figures}). The last recording contains
    segments of corrupted data. These artefacts strongly influence
    scaling, fractal and nonlinear measures (see
    Figs.~\ref{fig:model_apnea} and \ref{fig:arti_slp}), and can lead
    to spurious differences between young and elderly subjects.  }
  \label{fig:excluded}
\end{figure}

\begin{figure}[htbp]
  \centering
  \includegraphics[width=\linewidth]{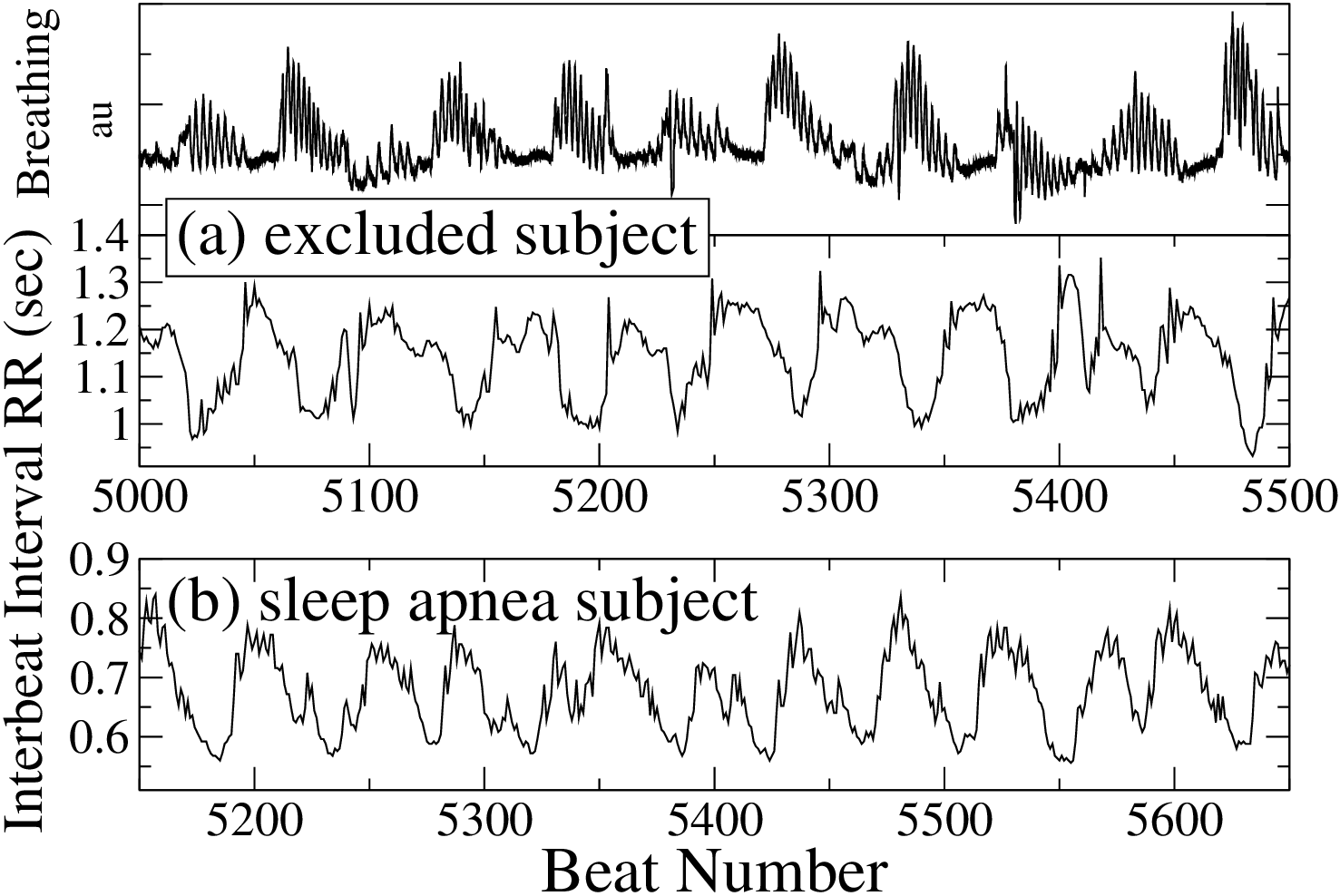}
  \label{fig:comparision_slp_excluded}
  \caption{
    Segments of interbeat RR interval time series for (a) an elderly
    subject from the Fantasia database excluded from this study (shown
    in Fig.~\ref{fig:excluded}, top panel), and
    (b) a subject diagnosed with sleep apnea from the apnea-ECG
    database~\cite{PhysioNet}. Both subjects show very similar and
    pronounced periodic patterns with a period of about
    \unit[50]{beats}, matching the periodic patterns in the breathing
    record (top panel (a)). These patterns strongly affect the scaling
    analysis as demonstrated in Figs.~\ref{fig:model_apnea} and
    \ref{fig:arti_slp}.}
  \label{fig:apnea_figures}
\end{figure}

Our findings indicate that scale-invariant correlation and nonlinear
properties do \emph{not} significantly change in healthy elderly
subjects compared to young subjects.  This is in contrast to some
earlier studies, based on the same Fantasia database (or on a subset
of it), which have reported loss of fractal organization in heartbeat
fluctuations --- a behavior resembling Brownian motion (random walk
process) with $\alpha=1.5$ at small scales and white noise with
$\alpha=0.5$ over large scales~\cite{Journal:Iyengar1996,
  Journal:Pikkujamsa1999Circulation}, as well as a significant loss of
nonlinearity~\cite{Journal:Vargas2003PRE} with healthy aging. A
possible reason for these different findings may be the presence of
artefacts in the data such as segments of corrupted recordings or
certain periodic patterns (Fig.~\ref{fig:excluded}).  These periodic
patterns strongly resemble episodes of sleep apnea, as shown in
Fig.~\ref{fig:apnea_figures} (Top panel). Indeed, sleep apnea may be
present in the elderly subjects from the Fantasia database, since they
have not been specifically screened for sleep apnea.  Further, ECG
recordings were taken when subjects were watching a calming movie for
\unit[2]{hours} in a semi-recumbent or supine posture during which
subjects may have fallen asleep for periods of time, when apnea
episodes are likely to occur.

\begin{figure*}[htbp]
  \centering{
  \subfigure
  {
    \includegraphics[width=0.45\linewidth]{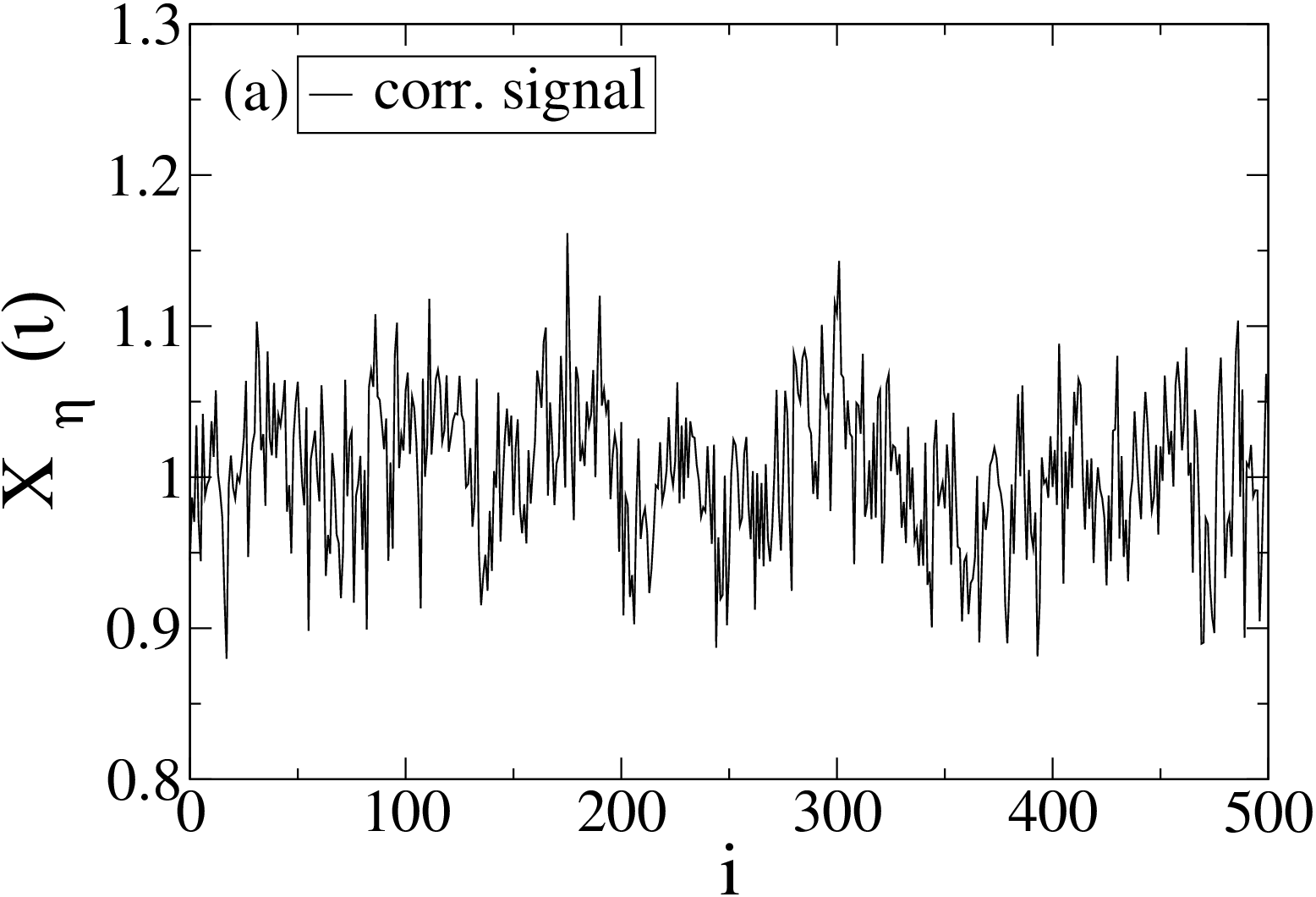}
    \label{fig:noise}
  }
  \subfigure
  {
    \includegraphics[width=0.45\linewidth]{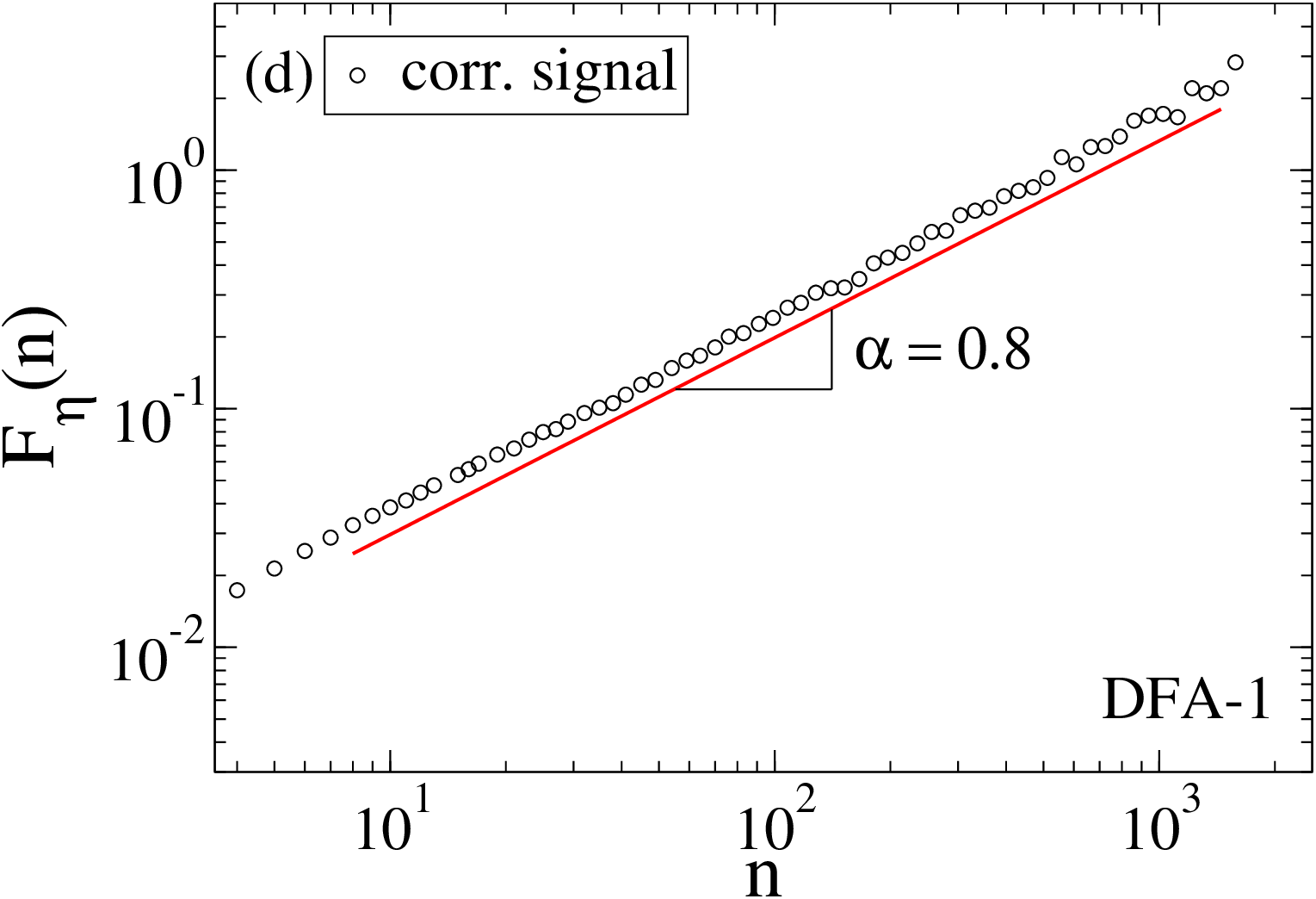}
    \label{fig:noise_dfa}
  }
  \\
  \subfigure
  {
    \includegraphics[width=0.45\linewidth]{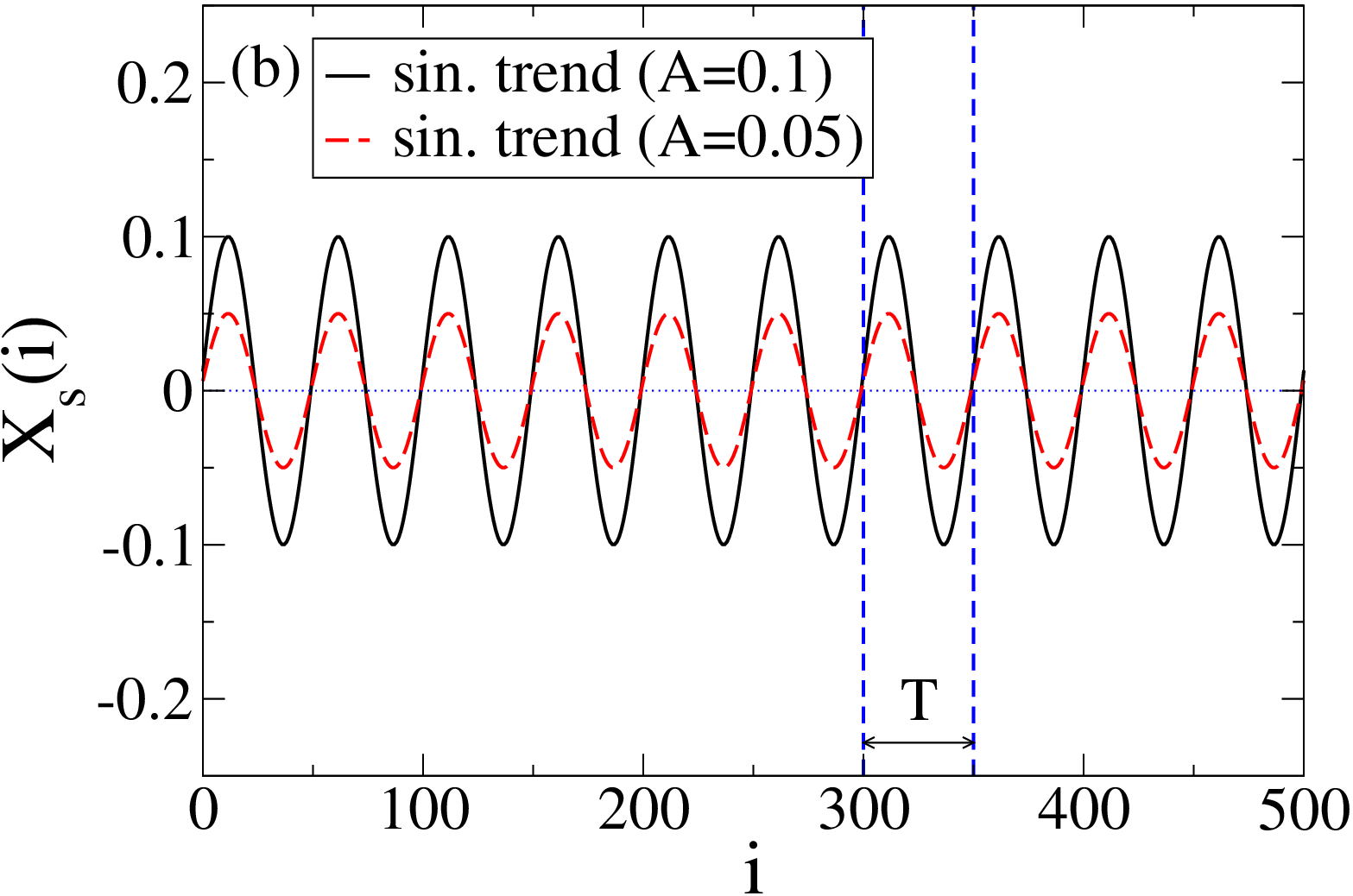}
    \label{fig:sin}
  }
  \subfigure
  {
    \includegraphics[width=0.45\linewidth]{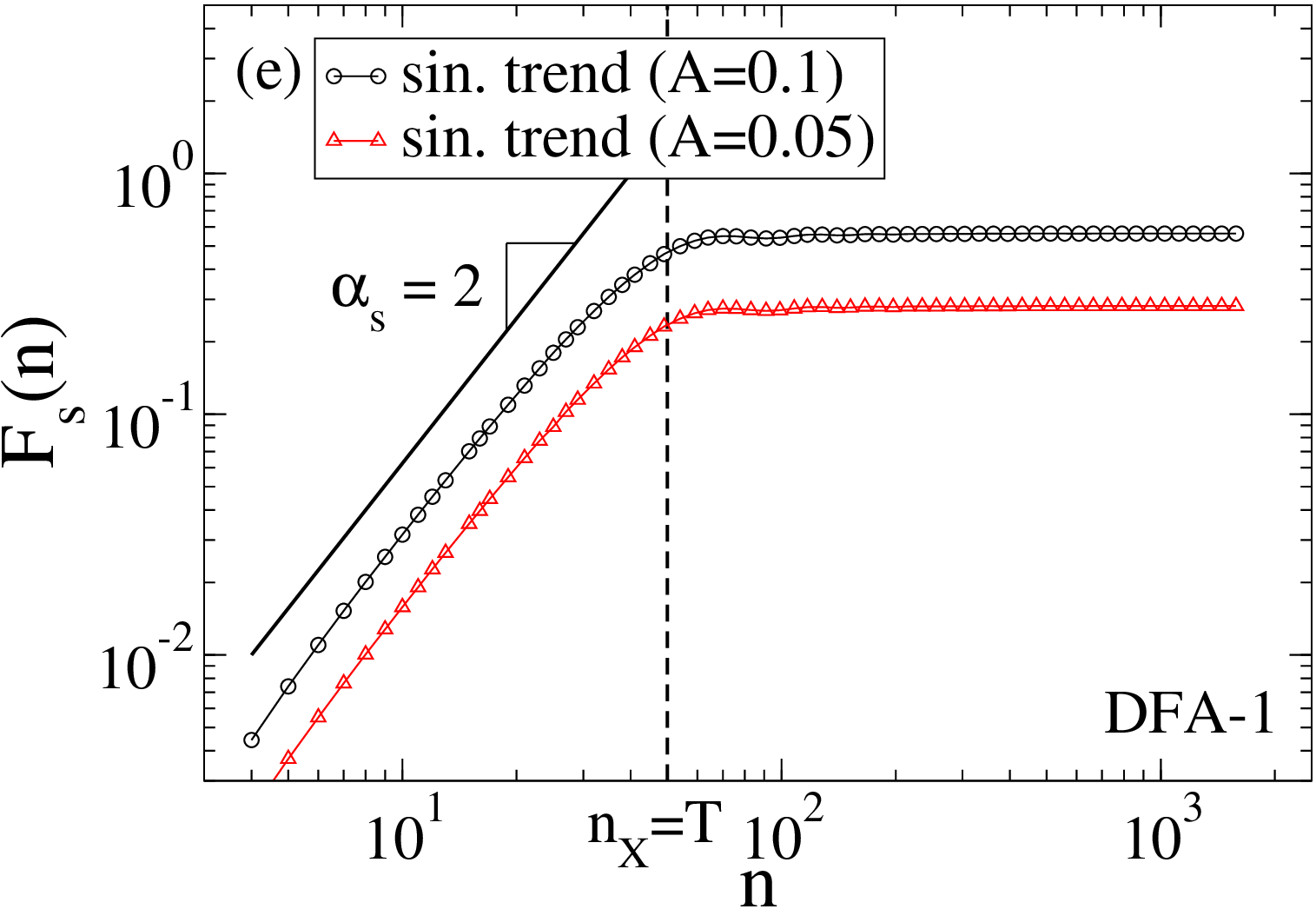}
    \label{fig:sin_dfa}
  }
  \\
  \subfigure
  {
    \includegraphics[width=0.45\linewidth]{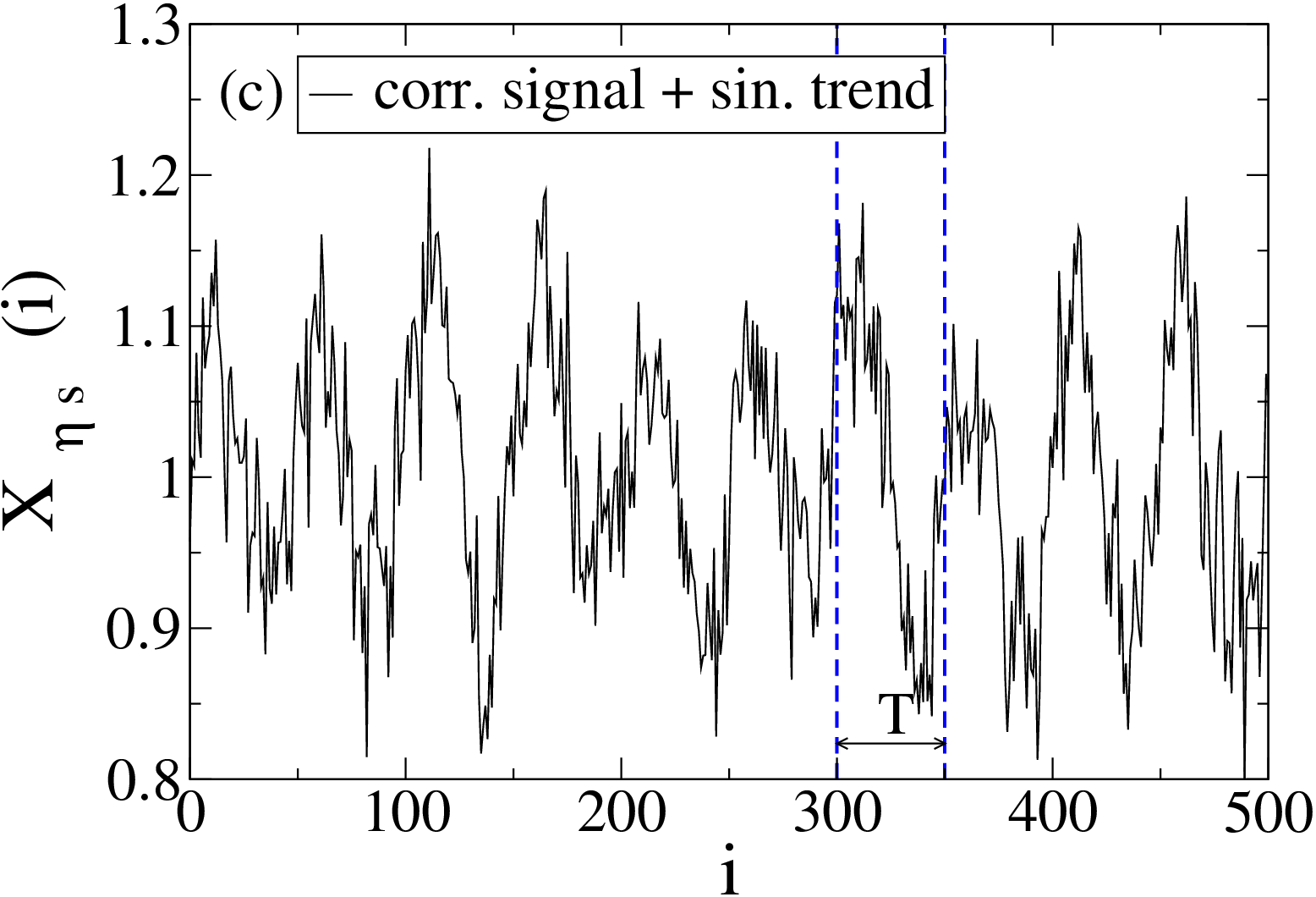}
    \label{fig:superpose_sin_noise}
  }
  \subfigure
  {
    \includegraphics[width=0.45\linewidth]{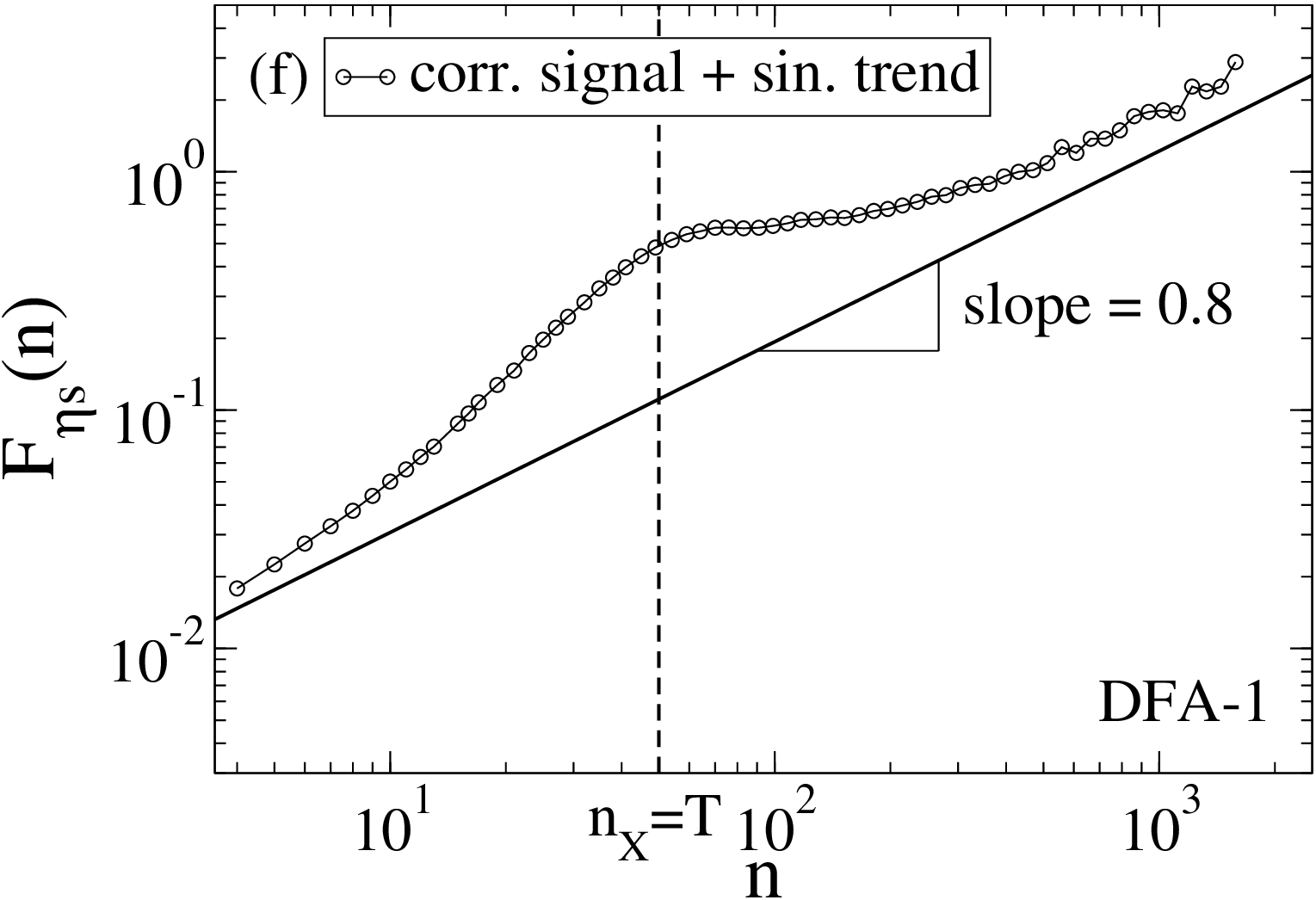}
    \label{fig:superpose_sin_noise_dfa}
  }
  }
  \caption{Modeling crossover behavior in the scaling of heartbeat
    dynamics associated with periodic patterns.  (a) Artificially
    generated fractal signal $X_\eta$ with long-range power-law
    correlations, average value and standard deviation as observed in
    healthy heartbeat data.  (b) A sinusoidal signal $X_s$ with period
    and amplitude matching the period $T$ and amplitude $A$ of typical
    sleep apnea patterns embedded in heartbeat interval time series as
    shown in Fig.~\ref{fig:apnea_figures}.  (c) Superposition of the
    signals $X_\eta$ in (a) and $X_s$ in (b). Note the apparent
    similarity between the signal $X_{\eta{}s}$ and the time series shown in
    Fig.~\ref{fig:apnea_figures}.  (d) Fluctuation function
    $F_{\eta}(n)$ obtained using DFA-$1$ for the signal $X_\eta$ in
    (a).  (e) Fluctuation function $F_{s}(n)$ obtained using DFA-$1$
    for the signal $X_s$ in (b).  The position of the crossover
    $n_{\times}$ corresponds to the period $T$ in $X_s$.  Changing $A$
    leads to a vertical shift of $F_{s}(n)$. (f) Fluctuation function
    $F_{\eta{}s}(n)$ obtained using DFA-$1$ for the signal
    $X_{\eta{}s}$ in (c).  Note the appearance of a kink with a
    crossover at $n_{\times} \approx T$ as observed in (e).}
  \label{fig:model_apnea}
\end{figure*}

\begin{figure}[htbp]
  \centering{
  \includegraphics[width=\linewidth]{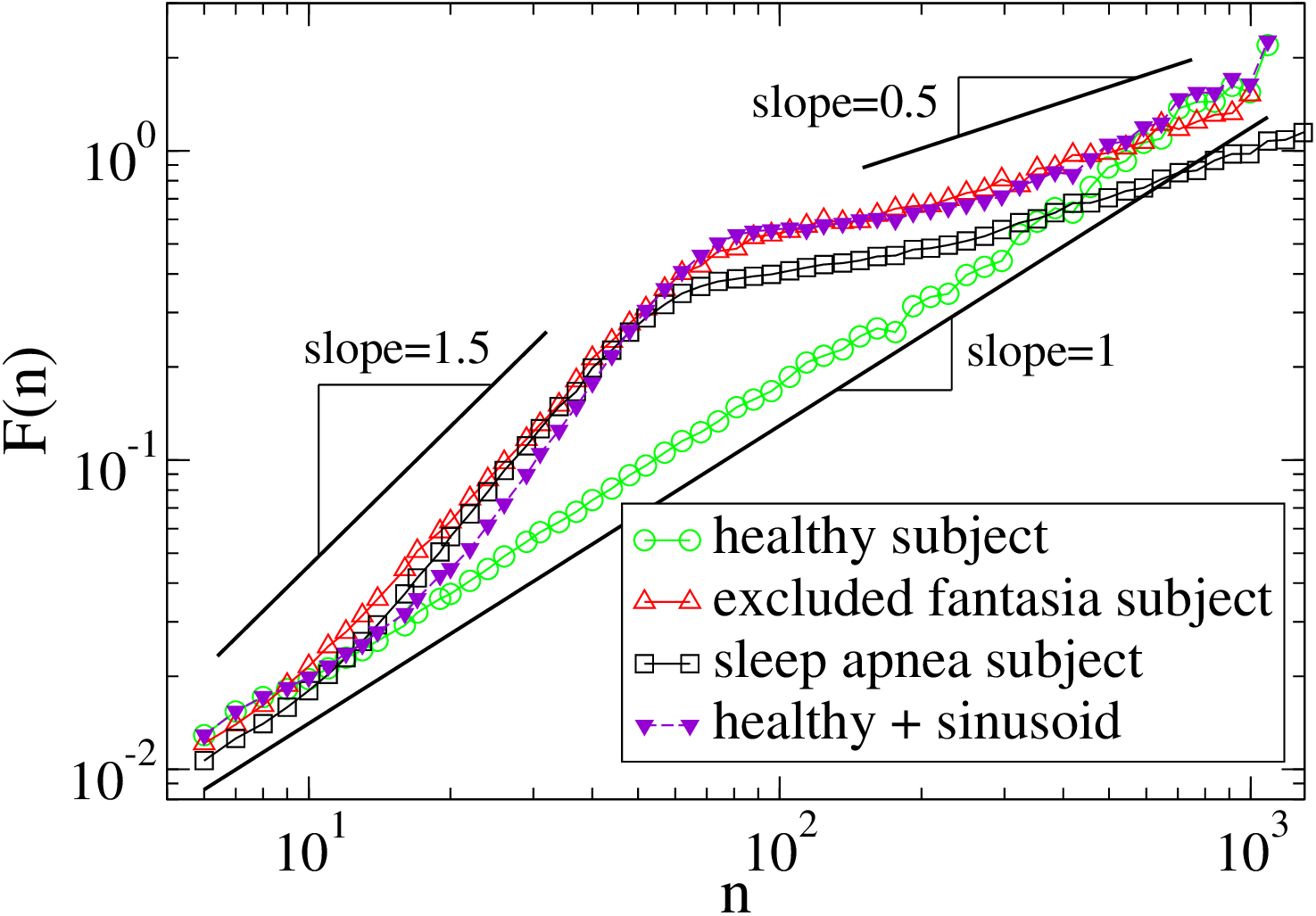}
  }
  \caption{Scaling functions $F(n)$ vs. time scale $n$ obtained for
    the heartbeat intervals using \mbox{DFA-$2$} for a healthy subject
    taken from (MIT-BIH Normal Sinus Rhythm Database~\cite{PhysioNet})
    ($\circ$), a Fantasia database subject we excluded from this
    study~\cite{PhysioNet} (shown in Fig.~\ref{fig:apnea_figures}(a)),
    a subject with diagnosed sleep apnea ($\Box$) (shown in
    Fig.~\ref{fig:apnea_figures}(b)), and a healthy subject with a
    superposed sinusoidal signal ($\nabla$). A period of
    $T=\unit[50]{beats}$ and an amplitude of $A=\unit[0.1]{sec}$ were
    chosen for the sinusoidal signal to model the effect of periodic
    patterns due to sleep apnea on the scaling function $F(n)$.  This
    effect leads to a change in the scaling exponent to 
    $\alpha\approx{}1.5$ (left of the crossover at $T$) and to
    $\alpha\approx{}0.5$ (right of the crossover), which may be
    the reason why earlier studies have reported loss of fractal
    organization in heartbeat fluctuations with healthy aging~\cite{Journal:Iyengar1996, Journal:Pikkujamsa1999Circulation}.}
  \label{fig:arti_slp}
\end{figure}

The periodic patterns we observe in wide segments of the interbeat
interval recordings shown in Fig.~\ref{fig:excluded} and
Fig.~\ref{fig:apnea_figures}(a) have a period of approximately $30$ to
$60$ seconds, typical for apnea episodes
(Fig.~\ref{fig:apnea_figures}(b))~\cite{Journal:Ivanov1996Nature,
  Journal:Ivanov1998PhysA, Journal:Quan1999Sleep}.  Similar apnea-like
patterns are also present in the breathing records of some Fantasia
subjects (Figs.~\ref{fig:excluded} and \ref{fig:apnea_figures}(a)).
These periodic patterns have a very strong effect on the scaling
analysis, as shown in earlier studies~\cite{Journal:Kun2001PRE},
leading to a pronounced crossover at the time scale corresponding to
the period of the patterns.  This crossover separates a regime of
apparent Brownian-motion-type behavior with $\alpha\approx{}1.5$ at
smaller scales from a second regime of apparent white noise behavior
$\alpha\approx{}0.5$ at larger scales (Figs.~\ref{fig:model_apnea}(f)
and \ref{fig:arti_slp}) --- a behavior which in earlier
studies~\cite{Journal:Iyengar1996, Journal:Pikkujamsa1999Circulation}
has spuriously been attributed to changes in the cardiac
neuroautonomic control due to aging.

To model the effect which periodic patterns of sleep apnea have on the
scaling properties of heartbeat intervals, we first generate a fractal
correlated signal $X_\eta$ using the Makse et. al.
algorithm~\cite{Journal:Makse1996Phys.Rev.E}.  To account for the
statistical properties observed in heartbeat intervals, we rescale the
signal to have the mean value $\left<X_{\eta}(i)\right>=1$, standard
deviation $\sigma_{X_{\eta}}=0.05$, and correlation scaling exponent
$\alpha_{X_{\eta}}=0.8$ (Fig.~\ref{fig:model_apnea}(a)), which match
the group mean $\left<RR\right>$, standard deviation
$\left<\sigma_{RR}\right>$ (Table~\ref{tab:allresults}), and scaling
exponent value $\left<\alpha_2\right>$ (Appendix, Fig.~\ref{fig:Fn_RR_all} (c,
d)) of the elderly subjects in the Fantasia database.  To model the
periodic influence of sleep apnea on the heartbeat intervals, we
generate a sinusoidal signal, $X_s(i)= A \sin{}\left(2\pi i/T\right)$,
with a period $T=50$ (similar to the average period of 50 heartbeats
in apnea patterns) and amplitude $A=0.1$ (as observed in apnea patterns)
(Fig.~\ref{fig:model_apnea}(b)), and we superpose the sinusoidal
signal $X_s$ with the fractal correlated signal $X_{\eta}(i)$ to
obtain $X_{\eta{}s}(i)=X_{\eta{}}(i)+X_{s}(i)$
(Fig.~\ref{fig:model_apnea}(c)). We note that $X_{\eta{}s}(i)$
strongly resembles the data shown in Fig.~\ref{fig:excluded} and
Fig.~\ref{fig:apnea_figures}.

Applying the DFA analysis to the fractal signal $X_{\eta}$ we obtain
the scaling function $F_{\eta}(n)$ with a slope of $0.8$ across all
scales --- in agreement with the scaling exponent $\alpha=0.8$ we have
found for healthy subjects (Fig.~\ref{fig:model_apnea}(d)).  For the
sinusoidal signal $X_s$ the scaling function $F_{s}(n)$ exhibits a
crossover at scale $n_{\times} \approx T$, corresponding to the period
of $X_s$.  For scales $n_{\times} < T$, the fluctuation function
$F_{s}(n)$ exhibits an apparent scaling, $F_{s}(n)\sim{}\frac{A}{T}\,
n^{\alpha_s}$, with an exponent $\alpha_s=2$.  For scales $n_{\times}
> T$, due to the periodic property of the sinusoidal signal $X_s$, the
fluctuation function $F_{s}(n)$ is constant and independent of the
scale $n$, i.e., $F_{s}(n)\sim{} A\, T\, n^{\alpha_s}$, where
$\alpha_s=0$. Thus, changing the amplitude $A$ leads to a vertical
shift in $F_{s}(n)$
(Fig.~\ref{fig:model_apnea}(e))~\cite{Journal:Kun2001PRE}.

Applying the DFA analysis to our model signal $X_{\eta{}s}$, we
observe that $F_{\eta{}s}(n)$ exhibits a very pronounced kink (not
present in $F_{\eta{}}(n)$) with a crossover at $n_{\times} \approx T$
due to the sinusoidal trend (Fig.~\ref{fig:model_apnea}(f)). The
behavior of $F_{\eta{}s}(n)$ around  the kink is very similar to
$F_{s}(n)$ around $n_{\times} \approx T$. At small scales $n_{\times}
< T$ and at large scales $n_{\times} > T$ the fluctuation function 
$F_{\eta{}s}(n)$ converges to the scaling behavior expected for
$F_{\eta{}}(n)$. Testing our model for signals $X_\eta$ with different
values for $\alpha$, we find that the position of the crossover
$n_{\times}$ for $F_{\eta{}s}(n)$ does not depend on $\alpha$. Thus,
this type of crossover behavior in the scaling for different subjects
depends only on the period $T$ of the periodic patterns embedded in
the heartbeat signals.

We find that our model in Fig.~\ref{fig:model_apnea}(f) reproduces
well the crossover behavior in $F(n)$ observed for the sleep apnea
subject (Apnea-ECG Database~\cite{PhysioNet}) shown in
Fig.~\ref{fig:apnea_figures}(b). Indeed, a very similar kink in $F(n)$
is observed at scale $n \approx{} 50$ beats for this apnea subject, as
shown in Fig.~\ref{fig:arti_slp}.  Moreover, we find that this
behavior is also closely followed (as shown in
Fig.~\ref{fig:arti_slp}) by the Fantasia subject in
Fig.~\ref{fig:apnea_figures}(a).  Adding the same sinusoidal trend to a
real heartbeat signal from a healthy subject (MIT-BIH Normal Sinus
Rhythm Database~\cite{PhysioNet}) also leads to a very similar kink in
$F(n)$ (Fig.~\ref{fig:arti_slp}).

As we demonstrate in Fig.~\ref{fig:arti_slp}, the excluded Fantasia
subject shown in Fig.~\ref{fig:apnea_figures}(a) exhibits a scaling
curve very similar to the curve obtained from a recording during sleep
from a subject diagnosed with sleep apnea (Apnea-ECG
Database~\cite{PhysioNet}). Further, our model reproduces well the
crossover in the scaling behavior of $F(n)$ and demonstrates that this
crossover is due to the superposition of healthy heart rate dynamics
and a sinusoidal trend with approximately the same period and
amplitude as the periodic apnea patterns shown in
Fig.~\ref{fig:apnea_figures}.  Our model reproduces also the scaling
curve $F(n)$ obtained for the elderly Fantasia subject excluded from
this study and shown in Fig.~\ref{fig:apnea_figures}(a), suggesting
that artifacts may have been the reason why earlier
studies~\cite{Journal:Iyengar1996, Journal:Pikkujamsa1999Circulation,
  Journal:Vargas2003PRE} have reported scaling differences in
heartbeat dynamics between young and elderly subjects.

Our modeling results confirm that the presence of pronounced
crossovers for some of the elderly subjects in the Fantasia database
are due to periodic patterns embedded in the heart rate which strongly
resemble sleep apnea episodes, and thus, cannot be attributed to
changes in the underlying mechanism of cardiac neuroautonomic
regulation associated with healthy aging.  Since apnea is more prominent in
elderly subjects, our modeling results (Figs.~\ref{fig:model_apnea}
and \ref{fig:arti_slp}) explain why earlier studies using the same
Fantasia database have reported higher values for the scaling exponent
$\alpha_1$ at small scales $n$ and lower values for $\alpha_2$ at large
scales $n$ for the elderly subjects compared to the group of young
subjects~\cite{Journal:Iyengar1996, Journal:Vargas2003PRE},
claiming changes in cardiac regulation with healthy aging.

\section{Discussion}
Our investigations demonstrate the presence of robust correlation,
fractal and nonlinear properties in cardiac dynamics of healthy
elderly subjects which remain surprisingly stable when compared to
healthy young subjects. Specifically, we find that key dynamical
characteristics such as the correlation scaling exponent of heartbeat
fluctuations, the scaling exponent of the magnitude and sign of
interbeat increments, and the nonlinear fractal dimension
measure do not significantly change with advanced age. Because the
scaling exponents $\alpha$ and the fractal dimension measure $D$
quantify a robust scale-invariant fractal and nonlinear structure in
heartbeat fluctuations~\cite{Journal:Peng1995Chaos,
  Journal:Ivanov1996Nature, Journal:Ivanov1999EPL,
  Journal:Ivanov1999Nature}, and have been shown to reflect underlying
mechanisms of cardiac control~\cite{Journal:Ivanov1998EurophysL,
  Journal:Ivanov2001Chaos, Journal:Amaral2001PRL,
  Journal:Goldberger2002PNAS}, our findings indicate that important
aspects of heartbeat regulation do not break down with healthy aging.
Moreover, we observe no significant change in these scaling and
nonlinear measures when comparing healthy elderly subjects with the
same subjects recorded five years later.

These findings do not support the hypothesis that healthy aging may be
associated with such a change in the mechanism of cardiac
neuroautonomic control that would lead to a loss of all aspects of
physiologic complexity. In contrast, we find that fundamental
scale-invariant and nonlinear properties of heartbeat dynamics remain
unchanged. Further, our findings do not support the hypothesis of a
gradual change of cardiac dynamics under healthy conditions with
advanced age, as key properties of these dynamics, including heart
rate variability (Table~\ref{tab:allresults}), remain stable in
healthy elderly subjects with advancing age.  Indeed, in agreement
with previous studies~\cite{Journal:Tsuji1994Circulation,
  Journal:Iyengar1996, Journal:Pikkujamsa1999Circulation,
  Journal:Corino2006}, we find a significant reduction in heart rate
variability as measured by $\sigma_{RR}$ (SDNN) and
$\sigma_{\Delta{}RR}$ (RMSSD) (although not in the average heart rate)
in healthy elderly subjects compared to healthy young subjects
(Table~\ref{tab:allresults}). The observed reduction in heart rate
variability is also in agreement with decrease of the commonly used
Approximate Entropy (ApEn) measure with aging, as reported
earlier~\cite{Journal:Corino2006}, and often interpreted as loss of
complexity. However, comparing elderly subjects with the same subjects
years later we do not find a further reduction in interbeat
variability. Moreover, we do not observe a loss in the scale-invariant
fractal and nonlinear features in healthy elderly compared to healthy
young subjects, indicating that the process of aging, even in elderly
healthy subjects, may not result in a gradual change of the mechanism
of control. Our findings support the hypothesis that (i) only certain
aspects of cardiac regulation may change with advanced age: these
aspects are related to decreased responsiveness to external and
internal stimuli, leading to reduced heart rate variability; (ii) other
fundamental features of the neuroautonomic cardiac control may remain
stable and unchanged with healthy aging: These features are related to
the network of nonlinear feedback loops responsible for the
neuroautonomic regulation at different time scales, leading to
scale-invariant cascades in heartbeat
fluctuations~\cite{Journal:Ivanov1998EurophysL, Journal:Lin2001,
  Journal:Ivanov2001Chaos}.

This new emerging picture of healthy aging is fundamentally different
from the changes in neural regulation of cardiac dynamics under
pathologic conditions~\cite{Journal:Ho1997Circulation,
  Journal:Maekikallio1997AJC, Laitio2000,
  Journal:Huikuri2000Circulation}, and also differs from previous
studies reporting breakdown of the scale-invariant and nonlinear
features of heartbeat dynamics in
elderly~\cite{Journal:Lipsitz1992Jama, Journal:Iyengar1996,
  Journal:Goldberger2002Commentary, Journal:Vargas2003PRE}. Indeed,
suppression of parasympathetic tone and dominance of sympathetic
inputs, typical for subjects with congestive heart failure, lead to
changes in cardiac dynamics associated with higher heart
rate~\cite{Journal:Saul1988AJC, Journal:Yamamoto1995}, lower heart
rate variability~\cite{Journal:Wolf1978}, relative loss of the
scale-invariant long-range correlations in the heartbeat fluctuations
with scaling exponent $\alpha$ between $1.25$ and $1.4$ (closer to
$\alpha=1.5$ corresponding to Brownian motion, i.e., random
walk)~\cite{Journal:Peng1995Chaos}, reduced 
responsiveness~\cite{Journal:Bernaola2001PRL}, as well as breakdown of
nonlinearity and multifractality~\cite{Journal:Poon1997Nature,
  Journal:Ivanov1999Nature, Journal:Ivanov2001Chaos,
  Journal:Amaral2001PRL}.  In contrast to such pathologic
perturbations, healthy aging appears to be accompanied only by a
reduction in heart rate variability as measured by $\sigma_{RR}$ and
$\sigma_{\Delta{}RR}$, while the heart rate, the scaling and nonlinear
properties remain on average unchanged. This important dissociation
between heart rate variability on one side and the scale-invariant and
nonlinear temporal organization of heartbeat fluctuations on the other
side may be specific for the process of aging, and suggests that the
alterations in the cardiac control mechanism with advanced age differ
conceptually from the mechanistic changes in the autonomic regulation
associated with pathologic conditions. More specifically, the reduced
heart rate variability with advanced age suggests a reduced 
responsiveness of cardiac control to external and internal stimuli,
and thus a reduced strength of feedback interactions. However, the
cascade of nonlinear feedback loops~\cite{Journal:Ivanov1998EurophysL,
  Journal:Lin2001, Journal:Ivanov2001Chaos} controlling the dynamics
across different time scales may remain intact in healthy elderly
subjects without breaking down at a particular scale or across a range
of scales, as the scale-invariant fractal and nonlinear properties
appear to remain stable with advanced age
(Table~\ref{tab:allresults}).  This is not the case with pathologic
conditions such as congestive heart failure, where the
self-organization of neural feedback interactions indeed breaks down
across time scales, shifting the dynamics closer to a process which is
more random (loss of long-range power-law correlations) and is closer
to a linear process (loss of nonlinearity and multifractality).

The value of the correlation exponent $\alpha_2\approx{}0.8$ we
observe at intermediate and large time scales for both young and
elderly Fantasia subjects (Figs.~\ref{fig:DFA_F_n_all_rep} and
\ref{fig:Fn_rr_stein_rep}) is consistent with earlier reports of a
very similar value of $\alpha_2\approx{}0.85$ for healthy subjects
during sleep, compared to $\alpha\approx{}1$ for the same subjects
during wake and daily activity~\cite{Journal:Ivanov1999EPL}. This is
also in agreement with studies of heartbeat dynamics of healthy
subjects during rest and exercise, with $\alpha\approx{}0.8$ for rest
and $\alpha\approx{}1.1$ during exercise~\cite{Journal:Karasik2002PRE,
  Journal:Martinis2004, Journal:Echeverria2006}. Indeed, the Fantasia
subjects were recorded under conditions of rest (see Data and Methods
Section~\ref{sec:data})~\cite{PhysioNet}.  Our findings of
$\alpha\approx{}0.8$ consistently for both healthy young and healthy
elderly subjects from the Fantasia database are further supported by
our analysis of data from the longitudinal SHHS study, where the same
elderly subjects were recorded during sleep several year later. These
observations of $\alpha<1$ are not due to artefacts in the heartbeat
time series related to sleep apnea, as full polysomnographic data were
recorded for the SHHS subjects indicating the apnea episodes, and we
have excluded the apnea segments in the data from our analysis.
Moreover, our preliminary results (a focus of a subsequent study)
indicate no significant differences between young and elderly subjects
even when we account for REM and NREM sleep stages.  Since there is no
statistically significant difference in the value of the scaling
exponent $\alpha$ between the young and elderly subjects from both
databases, the $\alpha$-value lower than $1$ is not likely to be
related to a mechanistic breakdown of cardiac control with advanced
age as previously suggested~\cite{Journal:Iyengar1996,
  Journal:Pikkujamsa1999Circulation}. Rather, this decrease in
$\alpha$ is most likely to be related to the normal regime of cardiac
regulation during rest and sleep when parasympathetic tone dominates
during NREM sleep stages, leading to stronger anti-correlations with
$\alpha\approx{}0.8$ in the heartbeat
fluctuations~\cite{Journal:Ivanov1998EurophysL, Journal:Ivanov1999EPL,
  Journal:Bunde2000PRL, Journal:Karasik2002PRE,
  Journal:Kantelhardt2003EPL}.

We find very similar results for the scaling exponent
$\alpha^{\mathrm{mag}}$ for the magnitude of the interbeat increments
between young and elderly subjects in the Fantasia database
(Table~\ref{tab:allresults}), as well as between the young Fantasia
subjects and the elderly subjects from the SHHS database (see
$p$-values reported in Results Section~\ref{sec:results_MSA}).  These
findings do \emph{not} support the hypotheses that the nonlinear
properties, as measured by the magnitude scaling exponent
$\alpha^{\mathrm{mag}}$ and encoded in the Fourier
phases~\cite{Journal:Theiler1992PhysicaD}, are lost with advanced age
in healthy subjects under resting conditions.  We note that our
results for the magnitude exponents for the young and elderly subjects
from both databases are in agreement with previous studies reporting
nonlinear magnitude correlations in healthy heartbeat
dynamics~\cite{Journal:Ashkenazy2001PRL}, and more specifically with
the magnitude exponent values found in the heartbeat fluctuations of
healthy subjects during sleep~\cite{Journal:Kantelhardt2002PRE,
  Journal:Kantelhardt2003EPL}.

Further, as the dynamics of the sign (directionality) of the interbeat
increments is directly related to inputs of the sympathetic and
parasympathetic branches of the autonomic nervous system modulating
the heart rate in opposite directions, our findings of similar scaling
for the sign series for both young and elderly healthy subjects
(Table~\ref{tab:allresults}) indicate that fundamental features of the
cardiac control mechanism remain unchanged with advanced age.  We also
note that our results for the sign scaling exponent
$\alpha^{\mathrm{sgn}}$ for the young and elderly subjects from both
databases are in agreement with the values reported in previous
studies for healthy subjects during rest~\cite{Journal:Karasik2002PRE}
and sleep~\cite{Journal:Kantelhardt2002PRE}.

While our results do not show a significant difference in the scaling and
nonlinear properties of heartbeat dynamics between healthy young and
healthy elderly subjects during rest and sleep, we note that under
conditions of high levels of physical activity and stress, which are
associated with a different regime of the neuroautonomic control,
these properties may differ between young and elderly subjects.

In summary, the observations reported here do not support the
hypothesis of a continuous gradual loss of the scaling and nonlinear
properties of cardiac dynamics with advanced age under healthy
conditions, as we do not find a statistically significant change in
these properties between the young and elderly subjects from the
Fantasia and the SHHS databases, as well as for the elderly subjects
from the SHHS database and the same subjects recorded five years
later. While cardiac dynamics in healthy elderly subjects is
characterized by markedly reduced variability compared to healthy
young subjects, the stability we observe in key fractal and nonlinear
characteristics with advanced age does not support the mechanistic
view of a breakdown of specific feedback loops at given time scales in
the neuroautonomic regulation (which would lead to appearance of
dominant time scales in the dynamics), or of a breakdown of the
feedback interactions in cardiac control across multiple time scales
(which would lead to random-like behavior in the dynamics). Indeed,
both dominant time scales and close-to-random behavior in cardiac
dynamics, have been observed under various pathologic conditions. In
contrast, cardiac dynamics under healthy aging appears not to belong
to this class of processes. Instead, our results indicate that the
inherent structure and temporal organization in the cascades of
nonlinear feedback loops underlying the cardiac neuroautonomic
regulation remains intact in healthy elderly subjects, thus preserving
the fractal and nonlinear features in heartbeat dynamics across all
time scales. The coupling strength of these neuronal feedback
interactions, however, is likely to diminish with advanced age,
leading to the observed reduction in heart rate variability and
dampened reflexive-type responsiveness in elderly compared to young
healthy subjects.

\section{Acknowledgments}
  This work was supported by NIH Grant No. 2 RO1 HL071972 and the
  Volkswagen Foundation.  We thank George B. Moody for discussions and
  help with the ARISTOTLE ECG-annotator.

  The findings in this report were based on data
  publicly available through the Sleep Heart Health Study (SHHS). However,
  the analyses and interpretation were not reviewed by members of the
  SHHS and do not reflect their approval.


\newpage{}
\appendix{}
\section{}
Results of our DFA and MSA analyses for the heartbeat interval
recordings for all young and elderly subjects in the Fantasia
database.  

\begin{figure}[htbp]
  \centering
  \subfigure{
    \includegraphics[width=\linewidth]{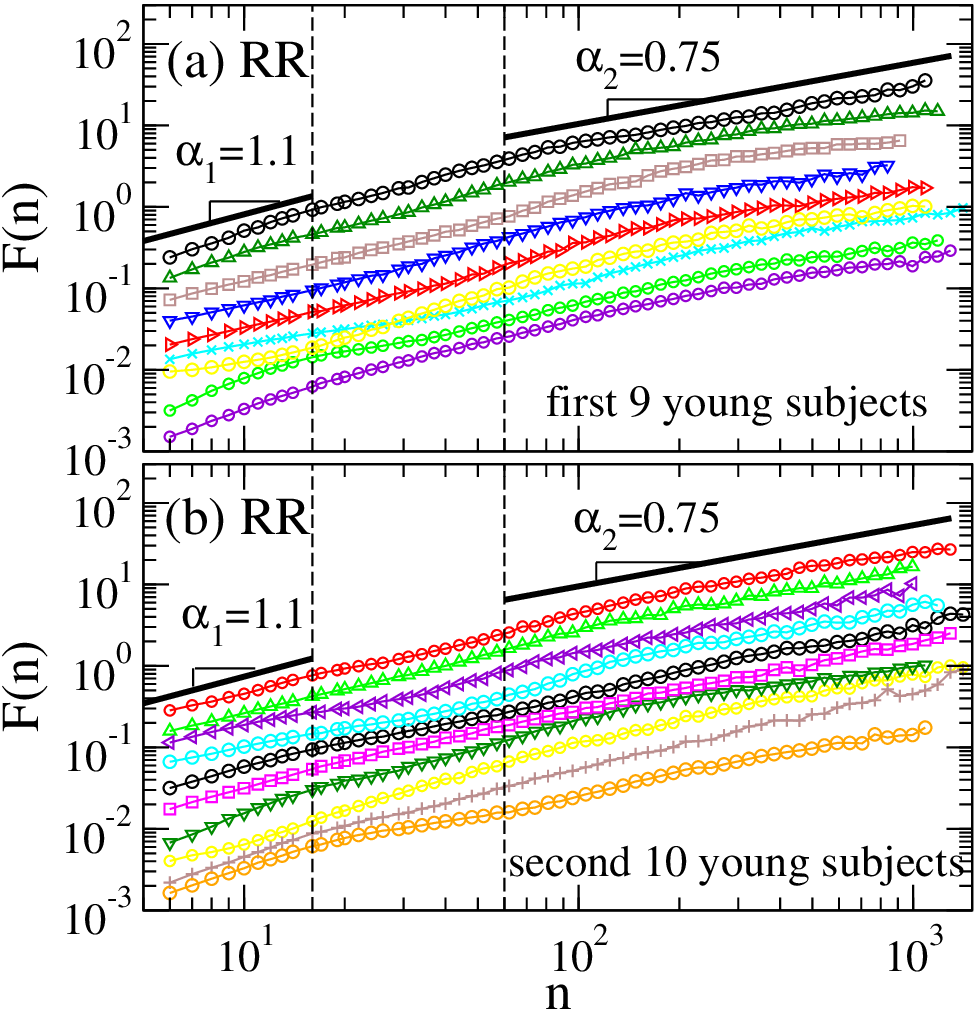}
    \label{fig:Fn_RR_young}
  }\\
  \vspace{-0.2cm}
  \subfigure{
    \includegraphics[width=\linewidth]{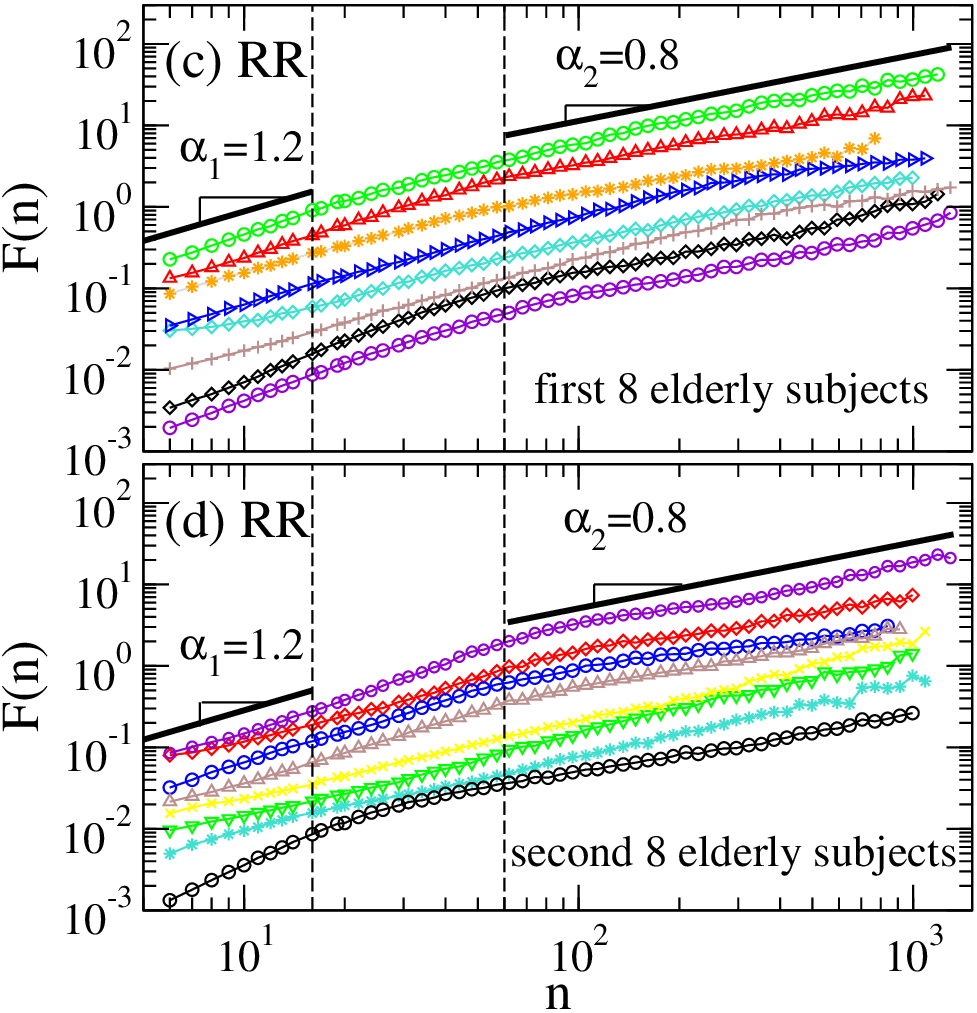}
    \label{fig:Fn_RR_elderly}
  }
  \caption{Scaling curves $F(n)$ versus time scale $n$ (in beat
    numbers) obtained for the RR heartbeat intervals using \mbox{DFA-$2$} for
    (a--b) 19 young healthy subjects and (c--d) 16 elderly healthy
    subjects in the Fantasia database. Despite certain inter-subject
    variability, there is a very common scaling behavior with a
    crossover from a higher average slope $\alpha_1$ at small time
    scales to a lower average slope $\alpha_2$ at large scales as
    represented by the solid lines and consistent with Fig.
    \ref{fig:DFA_F_n_all_rep} and Fig.~\ref{fig:Fn_rr_stein_rep}.
    Individual curves are vertically shifted to aid visual comparison.
    Group average statistics are presented in Table~\ref{tab:scaling_exp_DFA-2}. Vertical dashed lines indicate the
    range of fit.}
  \label{fig:Fn_RR_all}
\end{figure}

\noindent{}
All subjects show a consistent behavior with:

1) A smooth crossover from $\alpha_1\approx{}1.1$ at small time scales
to $\alpha_2\approx{}0.8$ at large scales for the heartbeat intervals
$RR$ for both the young and the elderly group (Fig.~\ref{fig:Fn_RR_all}).

\begin{figure}[htbp]
  \centering 
   \subfigure{
     \includegraphics[width=\linewidth]{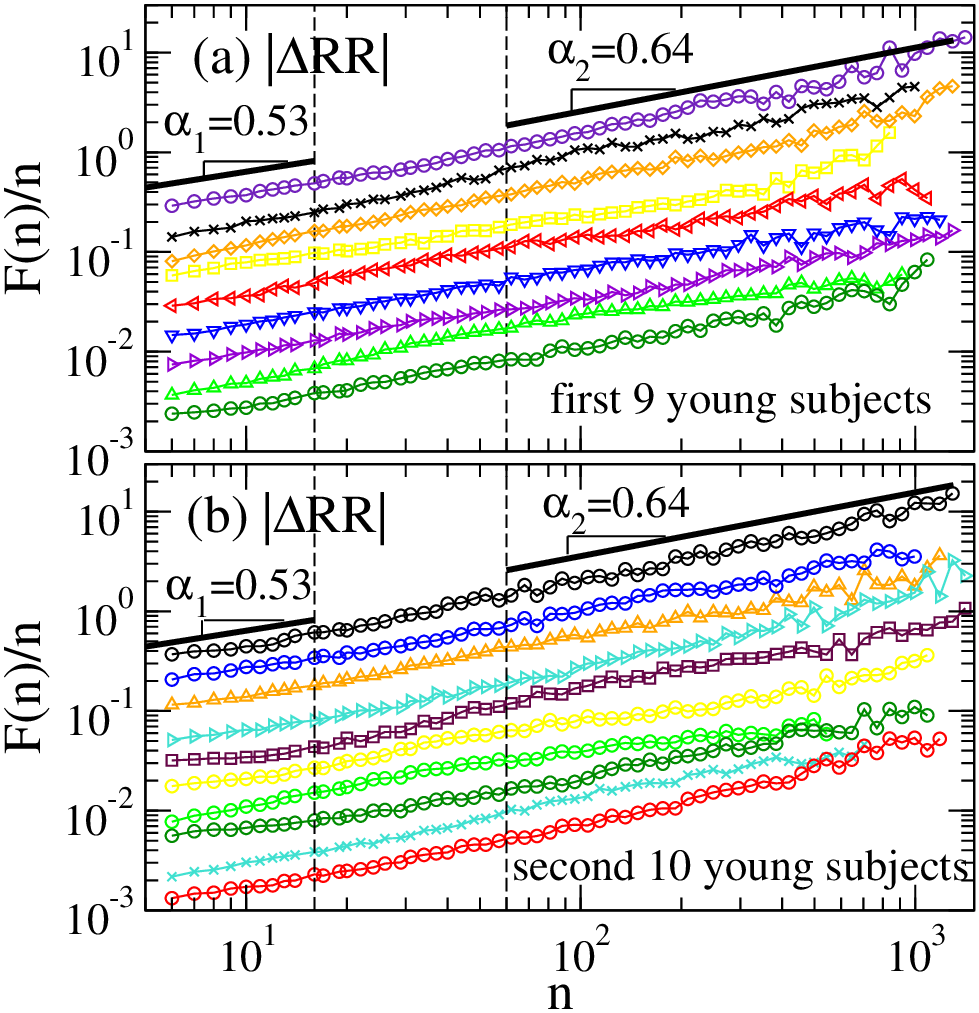}
     \label{fig:Fn_abs_young}
   }\\
   \vspace{-0.2cm}
   \subfigure{
     \includegraphics[width=\linewidth]{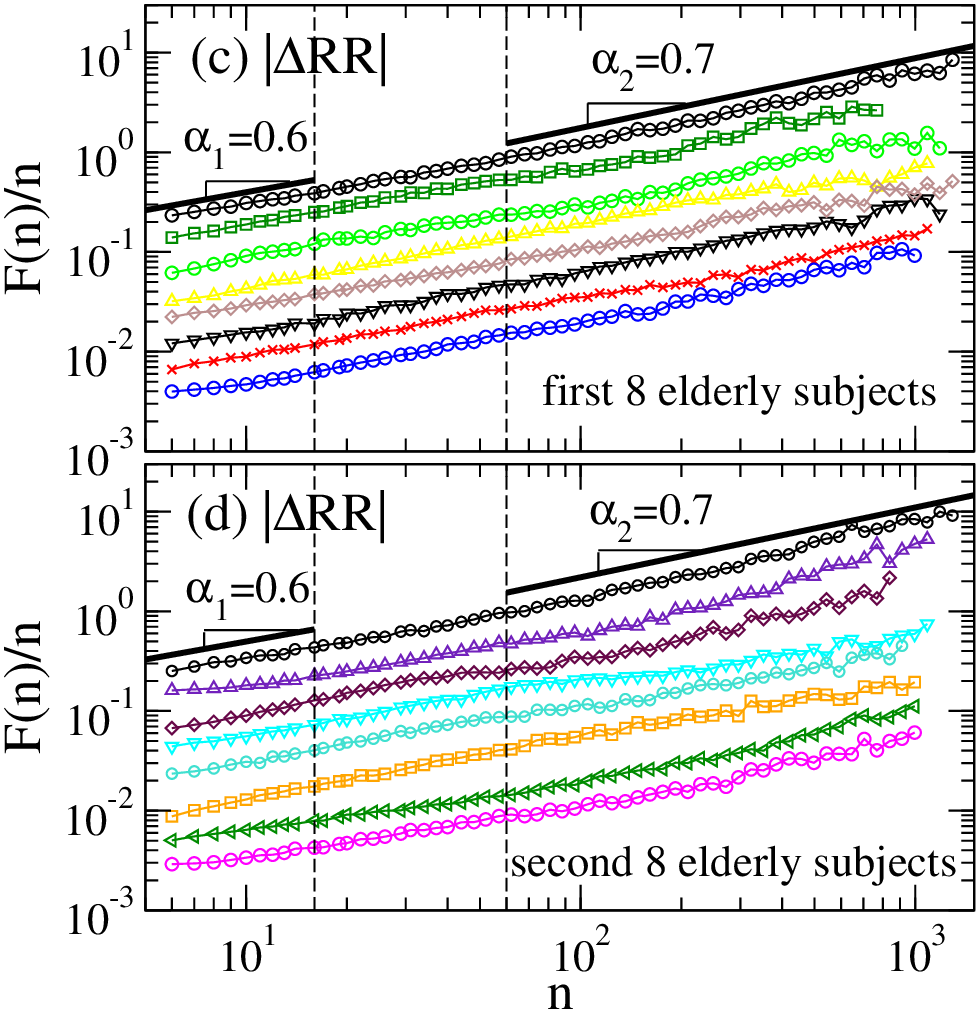}
     \label{fig:Fn_abs_old}
   }
   \caption{Scaling curves $F(n)$ versus time scale $n$ (in beat
    numbers) obtained for the magnitude of the
     interbeat increments $|\Delta{}RR|$ using \mbox{DFA-$2$} for
     (a--b) 19 healthy young subjects and (c--d) 16 healthy elderly
     subjects in the Fantasia database. Despite certain inter-subject
     variability, there is a common scaling behavior characterized by
     a group average exponent $\alpha_2\approx{}0.7$ at large scales
     for all groups as represented by the solid lines, indicating
     presence of long-term nonlinear properties encoded in the Fourier
     phases of the heartbeat time series similar to those shown in
     Fig.~\ref{fig:Fn_abs_all}. Curves are vertically shifted for
     clarity.  Vertical dashed lines indicate the range of fit.}
   \label{fig:Fn_abs_young_old}
\end{figure}

2) A smooth crossover from $\alpha^{\mathrm{mag}}_1\approx{}0.6$ at
small time scales to $\alpha^{\mathrm{mag}}_2\approx{}0.7$ at large
scales for the magnitude of the interbeat increments $|\Delta{}RR|$
for both the young and the elderly group
(Fig.~\ref{fig:Fn_abs_young_old}).

\begin{figure}
  \centering 
  \subfigure{
    \includegraphics[width=\linewidth]{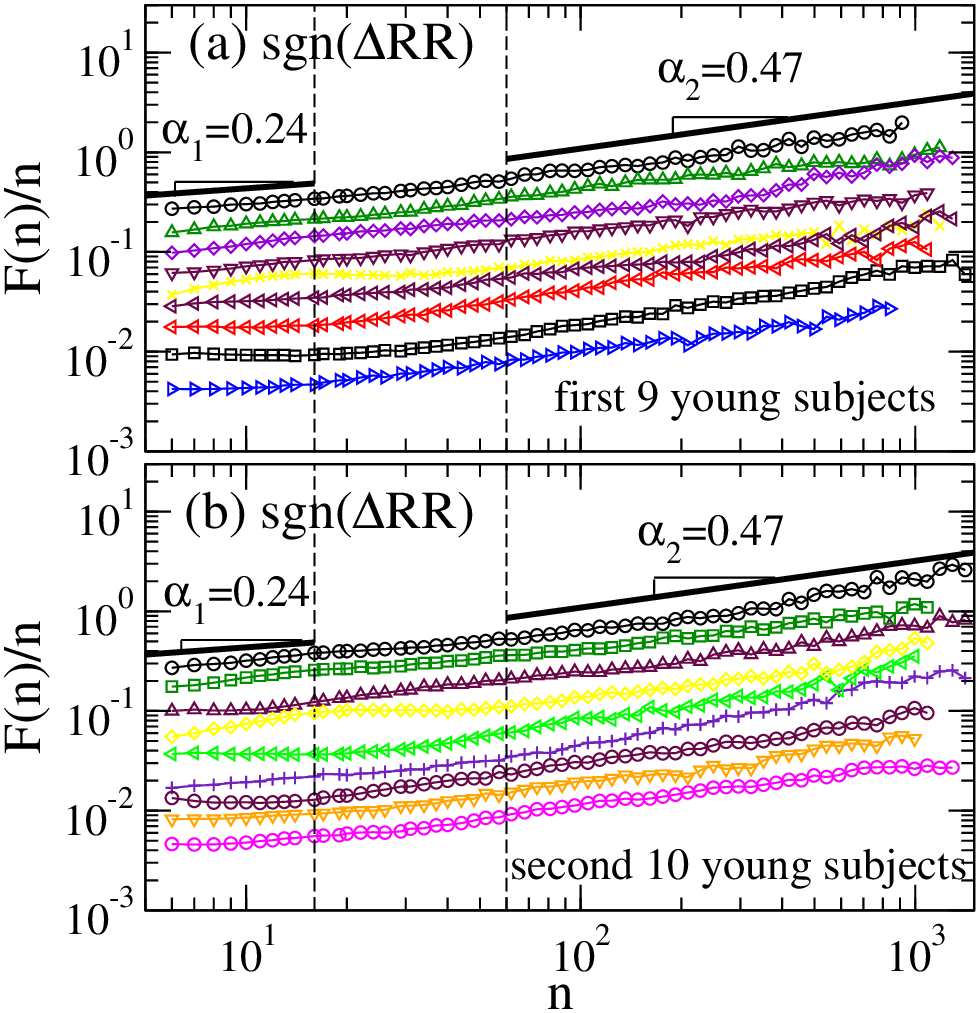}
    \label{fig:Fn_sgn_young}
  }\\
  \vspace{-0.2cm}
  \subfigure{
    \includegraphics[width=\linewidth]{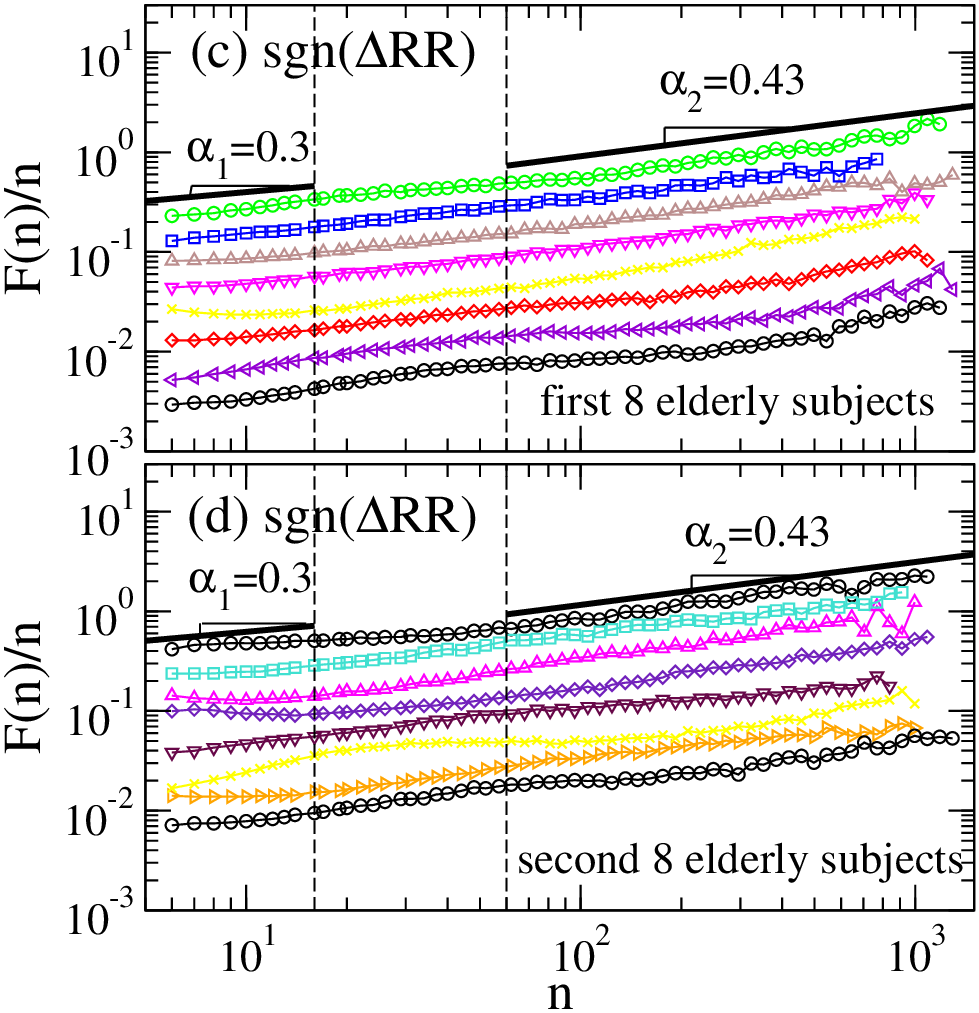}
    \label{fig:Fn_sgn_old}
  }
  \caption{Scaling curves $F(n)$ versus time scale $n$ (in beat
    numbers) obtained for the sign time series of the interbeat
    increments $\mathrm{sign}(\Delta{}RR)$ using \mbox{DFA-$2$} for
    (a--b) 19 healthy young subjects and (c--d) 16 healthy elderly
    subjects in the Fantasia database. All subjects exhibit a
    crossover from strongly (at small scales) to weakly (at large
    scales) anti-correlated behavior with no significant statistical
    difference between the young and elderly groups
    (Table~\ref{tab:scaling_exp_DFA-2}). Scaling curves are vertically
    shifted for clarity. Vertical dashed lines indicate the range of
    fit.}
  \label{fig:Fn_sgn_young_old_all}
\end{figure}

3) A crossover from $\alpha^{\mathrm{sgn}}_1\approx{}0.3$ at
small time scales to $\alpha^{\mathrm{sgn}}_2\approx{}0.45$ at large
scales for the sign of the interbeat increments $\mathrm{sign}(\Delta{}RR)$
for both the young and the elderly group
(Fig.~\ref{fig:Fn_sgn_young_old_all}).

The results show that these fractal correlation and nonlinear
properties of heartbeat dynamics do \emph{not} break down with healthy aging.

\end{document}